\documentclass[10pt,twoside,twocolumn]{IEEEtran}

\usepackage{cite}
\usepackage[T1]{fontenc}
\usepackage{graphicx}
\usepackage{amssymb}
\usepackage{amsmath}
\usepackage{amsthm}
\usepackage{subfigure}
\usepackage{booktabs} 
\usepackage{multirow}
\usepackage{microtype}
\usepackage{balance}
\usepackage{xcolor}
\usepackage[hidelinks]{hyperref}
\usepackage{mathdots}
\usepackage{epstopdf}
\usepackage{algorithm}
\usepackage{algpseudocode}
\usepackage{setspace}
\usepackage{footmisc}
\usepackage{tikz}
\usepackage{circledsteps}
\usepackage{mdframed}
\usepackage{balance}
\usepackage{bbm}
\usepackage{twemojis}

\usepackage{amssymb}
\usepackage{amsfonts}
\usepackage{mathrsfs}
\usepackage{xspace}
\usepackage{bm}
\usepackage{upgreek}

\newcommand{\safemath}[2]{\newcommand{#1}{\ensuremath{#2}\xspace}}

\safemath{\bma}{\mathbf{a}}
\safemath{\bmb}{\mathbf{b}}
\safemath{\bmc}{\mathbf{c}}
\safemath{\bmd}{\mathbf{d}}
\safemath{\bme}{\mathbf{e}}
\safemath{\bmf}{\mathbf{f}}
\safemath{\bmg}{\mathbf{g}}
\safemath{\bmh}{\mathbf{h}}
\safemath{\bmi}{\mathbf{i}}
\safemath{\bmj}{\mathbf{j}}
\safemath{\bmk}{\mathbf{k}}
\safemath{\bml}{\mathbf{l}}
\safemath{\bmm}{\mathbf{m}}
\safemath{\bmn}{\mathbf{n}}
\safemath{\bmo}{\mathbf{o}}
\safemath{\bmp}{\mathbf{p}}
\safemath{\bmq}{\mathbf{q}}
\safemath{\bmr}{\mathbf{r}}
\safemath{\bms}{\mathbf{s}}
\safemath{\bmt}{\mathbf{t}}
\safemath{\bmu}{\mathbf{u}}
\safemath{\bmv}{\mathbf{v}}
\safemath{\bmw}{\mathbf{w}}
\safemath{\bmx}{\mathbf{x}}
\safemath{\bmy}{\mathbf{y}}
\safemath{\bmz}{\mathbf{z}}
\safemath{\bmzero}{\mathbf{0}}
\safemath{\bmone}{\mathbf{1}}

\bmdefine{\biad}{a}
\bmdefine{\bibd}{b}
\bmdefine{\bicd}{c}
\bmdefine{\bidd}{d}
\bmdefine{\bied}{e}
\bmdefine{\bifd}{f}
\bmdefine{\bigd}{g}
\bmdefine{\bihd}{h}
\bmdefine{\biid}{i}
\bmdefine{\bijd}{j}
\bmdefine{\bikd}{k}
\bmdefine{\bild}{l}
\bmdefine{\bimd}{m}
\bmdefine{\bind}{n}
\bmdefine{\biod}{o}
\bmdefine{\bipd}{p}
\bmdefine{\biqd}{q}
\bmdefine{\bird}{r}
\bmdefine{\bisd}{s}
\bmdefine{\bitd}{t}
\bmdefine{\biud}{u}
\bmdefine{\bivd}{v}
\bmdefine{\biwd}{w}
\bmdefine{\bixd}{x}
\bmdefine{\biyd}{y}
\bmdefine{\bizd}{z}

\bmdefine{\bixid}{\xi}
\bmdefine{\bilambdad}{\lambda}
\bmdefine{\bimud}{\mu}
\bmdefine{\bithetad}{\theta}
\bmdefine{\biphid}{\phi}
\bmdefine{\bideltad}{\delta}

\safemath{\bmia}{\biad}
\safemath{\bmib}{\bibd}
\safemath{\bmic}{\bicd}
\safemath{\bmid}{\bidd}
\safemath{\bmie}{\bied}
\safemath{\bmif}{\bifd}
\safemath{\bmig}{\bigd}
\safemath{\bmih}{\bihd}
\safemath{\bmii}{\biid}
\safemath{\bmij}{\bijd}
\safemath{\bmik}{\bikd}
\safemath{\bmil}{\bild}
\safemath{\bmim}{\bimd}
\safemath{\bmin}{\bind}
\safemath{\bmio}{\biod}
\safemath{\bmip}{\bipd}
\safemath{\bmiq}{\biqd}
\safemath{\bmir}{\bird}
\safemath{\bmis}{\bisd}
\safemath{\bmit}{\bitd}
\safemath{\bmiu}{\biud}
\safemath{\bmiv}{\bivd}
\safemath{\bmiw}{\biwd}
\safemath{\bmix}{\bixd}
\safemath{\bmiy}{\biyd}
\safemath{\bmiz}{\bizd}

\safemath{\bmxi}{\bixid}
\safemath{\bmlambda}{\bilambdad}
\safemath{\bmmu}{\bimud}
\safemath{\bmtheta}{\bithetad}
\safemath{\bmphi}{\biphid}
\safemath{\bmdelta}{\bideltad}

\safemath{\bA}{\mathbf{A}}
\safemath{\bB}{\mathbf{B}}
\safemath{\bC}{\mathbf{C}}
\safemath{\bD}{\mathbf{D}}
\safemath{\bE}{\mathbf{E}}
\safemath{\bF}{\mathbf{F}}
\safemath{\bG}{\mathbf{G}}
\safemath{\bH}{\mathbf{H}}
\safemath{\bI}{\mathbf{I}}
\safemath{\bJ}{\mathbf{J}}
\safemath{\bK}{\mathbf{K}}
\safemath{\bL}{\mathbf{L}}
\safemath{\bM}{\mathbf{M}}
\safemath{\bN}{\mathbf{N}}
\safemath{\bO}{\mathbf{O}}
\safemath{\bP}{\mathbf{P}}
\safemath{\bQ}{\mathbf{Q}}
\safemath{\bR}{\mathbf{R}}
\safemath{\bS}{\mathbf{S}}
\safemath{\bT}{\mathbf{T}}
\safemath{\bU}{\mathbf{U}}
\safemath{\bV}{\mathbf{V}}
\safemath{\bW}{\mathbf{W}}
\safemath{\bX}{\mathbf{X}}
\safemath{\bY}{\mathbf{Y}}
\safemath{\bZ}{\mathbf{Z}}

\safemath{\bZero}{\mathbf{0}}
\safemath{\bOne}{\mathbf{1}}
\safemath{\bDelta}{\mathbf{\Delta}}
\safemath{\bLambda}{\mathbf{\UpLambda}}
\safemath{\bPhi}{\mathbf{\Upphi}}
\safemath{\bSigma}{\mathbf{\Upsigma}}
\safemath{\bOmega}{\mathbf{\Upomega}}
\safemath{\bTheta}{\mathbf{\Uptheta}}

\bmdefine{\biAd}{A}
\bmdefine{\biBd}{B}
\bmdefine{\biCd}{C}
\bmdefine{\biDd}{D}
\bmdefine{\biEd}{E}
\bmdefine{\biFd}{F}
\bmdefine{\biGd}{G}
\bmdefine{\biHd}{H}
\bmdefine{\biId}{I}
\bmdefine{\biJd}{J}
\bmdefine{\biKd}{K}
\bmdefine{\biLd}{L}
\bmdefine{\biMd}{M}
\bmdefine{\biOd}{N}
\bmdefine{\biPd}{O}
\bmdefine{\biQd}{P}
\bmdefine{\biRd}{R}
\bmdefine{\biSd}{S}
\bmdefine{\biTd}{T}
\bmdefine{\biUd}{U}
\bmdefine{\biVd}{V}
\bmdefine{\biWd}{W}
\bmdefine{\biXd}{X}
\bmdefine{\biYd}{Y}
\bmdefine{\biZd}{Z}

\bmdefine{\biDelta}{\Delta}
\bmdefine{\biLambda}{\Lambda}
\bmdefine{\biPhi}{\Phi}
\bmdefine{\biSigma}{\Sigma}
\bmdefine{\biOmega}{\Omega}
\bmdefine{\biTheta}{\Theta}

\safemath{\bimA}{\biAd}
\safemath{\bimB}{\biBd}
\safemath{\bimC}{\biCd}
\safemath{\bimD}{\biDd}
\safemath{\bimE}{\biEd}
\safemath{\bimF}{\biFd}
\safemath{\bimG}{\biGd}
\safemath{\bimH}{\biHd}
\safemath{\bimI}{\biId}
\safemath{\bimJ}{\biJd}
\safemath{\bimK}{\biKd}
\safemath{\bimL}{\biLd}
\safemath{\bimM}{\biMd}
\safemath{\bimN}{\biNd}
\safemath{\bimO}{\biOd}
\safemath{\bimP}{\biPd}
\safemath{\bimQ}{\biQd}
\safemath{\bimR}{\biRd}
\safemath{\bimS}{\biSd}
\safemath{\bimT}{\biTd}
\safemath{\bimU}{\biUd}
\safemath{\bimV}{\biVd}
\safemath{\bimW}{\biWd}
\safemath{\bimX}{\biXd}
\safemath{\bimY}{\biYd}
\safemath{\bimZ}{\biZd}

\safemath{\bimDelta}{\biDelta}
\safemath{\bimLambda}{\biLambda}
\safemath{\bimPhi}{\biPhi}
\safemath{\bimSigma}{\biSigma}
\safemath{\bimOmega}{\biOmega}
\safemath{\bimTheta}{\biTheta}

\safemath{\setA}{\mathcal{A}}
\safemath{\setB}{\mathcal{B}}
\safemath{\setC}{\mathcal{C}}
\safemath{\setD}{\mathcal{D}}
\safemath{\setE}{\mathcal{E}}
\safemath{\setF}{\mathcal{F}}
\safemath{\setG}{\mathcal{G}}
\safemath{\setH}{\mathcal{H}}
\safemath{\setI}{\mathcal{I}}
\safemath{\setJ}{\mathcal{J}}
\safemath{\setK}{\mathcal{K}}
\safemath{\setL}{\mathcal{L}}
\safemath{\setM}{\mathcal{M}}
\safemath{\setN}{\mathcal{N}}
\safemath{\setO}{\mathcal{O}}
\safemath{\setP}{\mathcal{P}}
\safemath{\setQ}{\mathcal{Q}}
\safemath{\setR}{\mathcal{R}}
\safemath{\setS}{\mathcal{S}}
\safemath{\setT}{\mathcal{T}}
\safemath{\setU}{\mathcal{U}}
\safemath{\setV}{\mathcal{V}}
\safemath{\setW}{\mathcal{W}}
\safemath{\setX}{\mathcal{X}}
\safemath{\setY}{\mathcal{Y}}
\safemath{\setZ}{\mathcal{Z}}
\safemath{\emptySet}{\varnothing}

\safemath{\colA}{\mathscr{A}}
\safemath{\colB}{\mathscr{B}}
\safemath{\colC}{\mathscr{C}}
\safemath{\colD}{\mathscr{D}}
\safemath{\colE}{\mathscr{E}}
\safemath{\colF}{\mathscr{F}}
\safemath{\colG}{\mathscr{G}}
\safemath{\colH}{\mathscr{H}}
\safemath{\colI}{\mathscr{I}}
\safemath{\colJ}{\mathscr{J}}
\safemath{\colK}{\mathscr{K}}
\safemath{\colL}{\mathscr{L}}
\safemath{\colM}{\mathscr{M}}
\safemath{\colN}{\mathscr{N}}
\safemath{\colO}{\mathscr{O}}
\safemath{\colP}{\mathscr{P}}
\safemath{\colQ}{\mathscr{Q}}
\safemath{\colR}{\mathscr{R}}
\safemath{\colS}{\mathscr{S}}
\safemath{\colT}{\mathscr{T}}
\safemath{\colU}{\mathscr{U}}
\safemath{\colV}{\mathscr{V}}
\safemath{\colW}{\mathscr{W}}
\safemath{\colX}{\mathscr{X}}
\safemath{\colY}{\mathscr{Y}}
\safemath{\colZ}{\mathscr{Z}}

\safemath{\opA}{\mathbb{A}}
\safemath{\opB}{\mathbb{B}}
\safemath{\opC}{\mathbb{C}}
\safemath{\opD}{\mathbb{D}}
\safemath{\opE}{\mathbb{E}}
\safemath{\opF}{\mathbb{F}}
\safemath{\opG}{\mathbb{G}}
\safemath{\opH}{\mathbb{H}}
\safemath{\opI}{\mathbb{I}}
\safemath{\opJ}{\mathbb{J}}
\safemath{\opK}{\mathbb{K}}
\safemath{\opL}{\mathbb{L}}
\safemath{\opM}{\mathbb{M}}
\safemath{\opN}{\mathbb{N}}
\safemath{\opO}{\mathbb{O}}
\safemath{\opP}{\mathbb{P}}
\safemath{\opQ}{\mathbb{Q}}
\safemath{\opR}{\mathbb{R}}
\safemath{\opS}{\mathbb{S}}
\safemath{\opT}{\mathbb{T}}
\safemath{\opU}{\mathbb{U}}
\safemath{\opV}{\mathbb{V}}
\safemath{\opW}{\mathbb{W}}
\safemath{\opX}{\mathbb{X}}
\safemath{\opY}{\mathbb{Y}}
\safemath{\opZ}{\mathbb{Z}}
\safemath{\opZero}{\mathbb{O}}
\safemath{\identityop}{\opI}

\safemath{\veca}{\bma}
\safemath{\vecb}{\bmb}
\safemath{\vecc}{\bmc}
\safemath{\vecd}{\bmd}
\safemath{\vece}{\bme}
\safemath{\vecf}{\bmf}
\safemath{\vecg}{\bmg}
\safemath{\vech}{\bmh}
\safemath{\veci}{\bmi}
\safemath{\vecj}{\bmj}
\safemath{\veck}{\bmk}
\safemath{\vecl}{\bml}
\safemath{\vecm}{\bmm}
\safemath{\vecn}{\bmn}
\safemath{\veco}{\bmo}
\safemath{\vecp}{\bmp}
\safemath{\vecq}{\bmq}
\safemath{\vecr}{\bmr}
\safemath{\vecs}{\bms}
\safemath{\vect}{\bmt}
\safemath{\vecu}{\bmu}
\safemath{\vecv}{\bmv}
\safemath{\vecw}{\bmw}
\safemath{\vecx}{\bmx}
\safemath{\vecy}{\bmy}
\safemath{\vecz}{\bmz}

\safemath{\veczero}{\bmzero}
\safemath{\vecone}{\bmone}
\safemath{\vecxi}{\bmxi}
\safemath{\veclambda}{\bmlambda}
\safemath{\vecmu}{\bmmu}
\safemath{\vectheta}{\bmtheta}
\safemath{\vecphi}{\bmphi}
\safemath{\vecdelta}{\bmdelta}

\safemath{\matA}{\bA}
\safemath{\matB}{\bB}
\safemath{\matC}{\bC}
\safemath{\matD}{\bD}
\safemath{\matE}{\bE}
\safemath{\matF}{\bF}
\safemath{\matG}{\bG}
\safemath{\matH}{\bH}
\safemath{\matI}{\bI}
\safemath{\matJ}{\bJ}
\safemath{\matK}{\bK}
\safemath{\matL}{\bL}
\safemath{\matM}{\bM}
\safemath{\matN}{\bN}
\safemath{\matO}{\bO}
\safemath{\matP}{\bP}
\safemath{\matQ}{\bQ}
\safemath{\matR}{\bR}
\safemath{\matS}{\bS}
\safemath{\matT}{\bT}
\safemath{\matU}{\bU}
\safemath{\matV}{\bV}
\safemath{\matW}{\bW}
\safemath{\matX}{\bX}
\safemath{\matY}{\bY}
\safemath{\matZ}{\bZ}
\safemath{\matzero}{\bmzero}

\safemath{\matDelta}{\bDelta}
\safemath{\matLambda}{\bLambda}
\safemath{\matPhi}{\bPhi}
\safemath{\matSigma}{\bSigma}
\safemath{\matOmega}{\bOmega}
\safemath{\matTheta}{\bTheta}

\safemath{\matidentity}{\matI}
\safemath{\matone}{\matO}

\safemath{\rnda}{A}
\safemath{\rndb}{B}
\safemath{\rndc}{C}
\safemath{\rndd}{D}
\safemath{\rnde}{E}
\safemath{\rndf}{F}
\safemath{\rndg}{G}
\safemath{\rndh}{H}
\safemath{\rndi}{I}
\safemath{\rndj}{J}
\safemath{\rndk}{K}
\safemath{\rndl}{L}
\safemath{\rndm}{M}
\safemath{\rndn}{N}
\safemath{\rndo}{O}
\safemath{\rndp}{P}
\safemath{\rndq}{Q}
\safemath{\rndr}{R}
\safemath{\rnds}{S}
\safemath{\rndt}{T}
\safemath{\rndu}{U}
\safemath{\rndv}{V}
\safemath{\rndw}{W}
\safemath{\rndx}{X}
\safemath{\rndy}{Y}
\safemath{\rndz}{Z}

\safemath{\rveca}{\bimA}
\safemath{\rvecb}{\bimB}
\safemath{\rvecc}{\bimC}
\safemath{\rvecd}{\bimD}
\safemath{\rvece}{\bimE}
\safemath{\rvecf}{\bimF}
\safemath{\rvecg}{\bimG}
\safemath{\rvech}{\bimH}
\safemath{\rveci}{\bimI}
\safemath{\rvecj}{\bimJ}
\safemath{\rveck}{\bimK}
\safemath{\rvecl}{\bimL}
\safemath{\rvecm}{\bimM}
\safemath{\rvecn}{\bimN}
\safemath{\rveco}{\bomO}
\safemath{\rvecp}{\bimP}
\safemath{\rvecq}{\bimQ}
\safemath{\rvecr}{\bimR}
\safemath{\rvecs}{\bimS}
\safemath{\rvect}{\bimT}
\safemath{\rvecu}{\bimU}
\safemath{\rvecv}{\bimV}
\safemath{\rvecw}{\bimW}
\safemath{\rvecx}{\bimX}
\safemath{\rvecy}{\bimY}
\safemath{\rvecz}{\bimZ}

\safemath{\rvecxi}{\bmxi}
\safemath{\rveclambda}{\bmlambda}
\safemath{\rvecmu}{\bmmu}
\safemath{\rvectheta}{\bmtheta}
\safemath{\rvecphi}{\bmphi}

\safemath{\rmatA}{\bimA}
\safemath{\rmatB}{\bimB}
\safemath{\rmatC}{\bimC}
\safemath{\rmatD}{\bimD}
\safemath{\rmatE}{\bimE}
\safemath{\rmatF}{\bimF}
\safemath{\rmatG}{\bimG}
\safemath{\rmatH}{\bimH}
\safemath{\rmatI}{\bimI}
\safemath{\rmatJ}{\bimJ}
\safemath{\rmatK}{\bimK}
\safemath{\rmatL}{\bimL}
\safemath{\rmatM}{\bimM}
\safemath{\rmatN}{\bimN}
\safemath{\rmatO}{\bimO}
\safemath{\rmatP}{\bimP}
\safemath{\rmatQ}{\bimQ}
\safemath{\rmatR}{\bimR}
\safemath{\rmatS}{\bimS}
\safemath{\rmatT}{\bimT}
\safemath{\rmatU}{\bimU}
\safemath{\rmatV}{\bimV}
\safemath{\rmatW}{\bimW}
\safemath{\rmatX}{\bimX}
\safemath{\rmatY}{\bimY}
\safemath{\rmatZ}{\bimZ}

\safemath{\rmatDelta}{\bimDelta}
\safemath{\rmatLambda}{\bimLambda}
\safemath{\rmatPhi}{\bimPhi}
\safemath{\rmatSigma}{\bimSigma}
\safemath{\rmatOmega}{\bimOmega}
\safemath{\rmatTheta}{\bimTheta}

\usepackage{amssymb}
\usepackage{amsfonts}
\usepackage{mathrsfs}
\usepackage{xspace}
\usepackage{bm}
\usepackage{fancyref}
\usepackage{textcomp}

\usepackage{multirow}
\usepackage{stmaryrd}

\newenvironment{textbmatrix}{	\setlength{\arraycolsep}{2.5pt}%
								\big[\begin{matrix}}{\end{matrix}\big]%
								\raisebox{0.08ex}{\vphantom{M}}}

\def\be{\begin{equation}}
\def\ee{\end{equation}}
\def\een{\nonumber \end{equation}}
\def\mat{\begin{bmatrix}}
\def\emat{\end{bmatrix}}
\def\btm{\begin{textbmatrix}}
\def\etm{\end{textbmatrix}}

\def\ba#1\ea{\begin{align}#1\end{align}}
\def\bas#1\eas{\begin{align*}#1\end{align*}}
\def\bs#1\es{\begin{split}#1\end{split}}
\def\bg#1\eg{\begin{gather}#1\end{gather}}
\def\bml#1\eml{\begin{multline}#1\end{multline}}
\def\bi#1\ei{\begin{itemize}#1\end{itemize}}

\newcommand{\lefto}{\mathopen{}\left}

\DeclareMathOperator*{\argmin}{arg\;min}		
\DeclareMathOperator*{\argmax}{arg\;max}		
\DeclareMathOperator{\Exop}{\opE}			

\newcommand{\orth}{\perp}					
\newcommand{\Ex}[2]{\ensuremath{\Exop_{#1}\lefto[#2\right]}} 	

\newcommand{\tp}[1]{\ensuremath{#1^{T}}} 		
\newcommand{\herm}[1]{\ensuremath{#1^{H}}} 	
\newcommand{\inv}[1]{\ensuremath{#1^{-1}}} 	
\newcommand{\pinv}[1]{\ensuremath{#1^{\dagger}}} 	

\safemath{\dirac}{\delta}					
\safemath{\krond}{\dirac}					

\safemath{\upto}{\uparrow}
\safemath{\downto}{\downarrow}
\safemath{\iu}{j}							
\safemath{\ev}{\lambda}						
\safemath{\hilseqspace}{l^{2}}				
\newcommand{\banachfunspace}[1]{\setL^{#1}}	
\safemath{\hilfunspace}{\banachfunspace{2}}	

\safemath{\SNR}{\textit{SNR}} 				
\safemath{\PAR}{\textit{PAR}} 				
\safemath{\No}{N_0}							
\safemath{\Es}{E_s}							
\safemath{\Eb}{E_b}							
\safemath{\EbNo}{\frac{\Eb}{\No}}
\safemath{\EsNo}{\frac{\Es}{\No}}

\DeclareMathOperator{\CHop}{\ensuremath{\opH}} 
\safemath{\tvir}{\rndh_{\CHop}}				
\safemath{\tvtf}{\rndl_{\CHop}}				
\safemath{\spf}{\rnds_{\CHop}}				
\safemath{\bff}{H_{\CHop}}					

\safemath{\ircf}{r_{h}}						
\safemath{\tftvcf}{r_{s}}					
\safemath{\tfcf}{r_{l}}						
\safemath{\bfcf}{r_{H}}						

\safemath{\tcorr}{c_h}						
\safemath{\scf}{c_{s}}						
\safemath{\tfcorr}{c_{l}}					
\safemath{\fcorr}{c_{H}}						

\safemath{\mi}{I}							
\safemath{\capacity}{C}						

\safemath{\normal}{\mathcal{N}}			
\safemath{\jpg}{\mathcal{CN}}			
\safemath{\mchain}{\leftrightarrow}		

\safemath{\dB}{\,\mathrm{dB}}
\safemath{\dBm}{\,\mathrm{dBm}}
\safemath{\Hz}{\,\mathrm{Hz}}
\safemath{\kHz}{\,\mathrm{kHz}}
\safemath{\MHz}{\,\mathrm{MHz}}
\safemath{\GHz}{\,\mathrm{GHz}}
\safemath{\s}{\,\mathrm{s}}
\safemath{\ms}{\,\mathrm{ms}}
\safemath{\mus}{\,\mathrm{\text{\textmu}s}}
\safemath{\ns}{\,\mathrm{ns}}
\safemath{\ps}{\,\mathrm{ps}}
\safemath{\meter}{\,\mathrm{m}}
\safemath{\mm}{\,\mathrm{mm}}
\safemath{\cm}{\,\mathrm{cm}}
\safemath{\m}{\,\mathrm{m}}
\safemath{\W}{\,\mathrm{W}}
\safemath{\mW}{\, \mathrm{mW}}
\safemath{\J}{\,\mathrm{J}}
\safemath{\K}{\,\mathrm{K}}
\safemath{\bit}{\,\mathrm{bit}}
\safemath{\nat}{\,\mathrm{nat}}

\safemath{\define}{\triangleq}			

\safemath{\equivalent}{\sim}
\safemath{\distas}{\sim}					
\safemath{\sdiff}{\Delta}				

\safemath{\reals}{\mathbb{R}}
\safemath{\positivereals}{\reals_{+}}
\safemath{\integers}{\mathbb{Z}}
\safemath{\posint}{\integers_{+}}
\safemath{\naturals}{\mathbb{N}}
\safemath{\posnaturals}{\naturals_{+}}
\safemath{\complexset}{\mathbb{C}}
\safemath{\rationals}{\mathbb{Q}}

\newcommand*{\fancyrefapplabelprefix}{app}		
\newcommand*{\fancyrefthmlabelprefix}{thm}		
\newcommand*{\fancyreflemlabelprefix}{lem}		
\newcommand*{\fancyrefcorlabelprefix}{cor}		
\newcommand*{\fancyrefdeflabelprefix}{def}		
\newcommand*{\fancyrefproplabelprefix}{prop}		
\newcommand*{\fancyrefexmpllabelprefix}{exmpl}
\newcommand*{\fancyrefalglabelprefix}{alg}		
\newcommand*{\fancyreftbllabelprefix}{tbl}		

\frefformat{vario}{\fancyrefseclabelprefix}{Sec.~#1}
\frefformat{vario}{\fancyrefthmlabelprefix}{Theorem~#1}
\frefformat{vario}{\fancyreftbllabelprefix}{Table~#1}
\frefformat{vario}{\fancyreflemlabelprefix}{Lemma~#1}
\frefformat{vario}{\fancyrefcorlabelprefix}{Corollary~#1}
\frefformat{vario}{\fancyrefdeflabelprefix}{Definition~#1}
\frefformat{vario}{\fancyreffiglabelprefix}{Fig.~#1}
\frefformat{vario}{\fancyrefapplabelprefix}{Appendix~#1}
\frefformat{vario}{\fancyrefeqlabelprefix}{(#1)}
\frefformat{vario}{\fancyrefproplabelprefix}{Proposition~#1}
\frefformat{vario}{\fancyrefexmpllabelprefix}{Example~#1}
\frefformat{vario}{\fancyrefalglabelprefix}{Algorithm~#1}

 \newtheorem{thm}{Theorem}

\safemath{\dictab}{[\,\dicta\,\,\dictb\,]}

\safemath{\ysig}{\bmy}
\safemath{\ysighat}{\hat{\ysig}}
\safemath{\ysigdim}{M}
\safemath{\xsig}{\bmx}
\safemath{\xsigdim}{N}
\safemath{\nx}{n_x}
\safemath{\zsig}{\bmz}
\safemath{\zsigdim}{\ysigdim}
\safemath{\rsig}{\bmr}
\safemath{\Adict}{\bA}
\safemath{\Adicttilde}{\widetilde{\Adict}}
\safemath{\Adictdim}{\outputdim\times\xsigdim}
\safemath{\avec}{\bma}
\safemath{\avectilde}{\tilde{\avec}}
\safemath{\Bdict}{\bB}
\safemath{\Bdicttilde}{\widetilde{\Bdict}}
\safemath{\Cdict}{\bC}
\safemath{\cvec}{\bmc}
\safemath{\Ddict}{\bD}
\safemath{\Ddictdim}{\ysigdim\times\xsigdim}
\safemath{\dvec}{\bmd}
\safemath{\Ddicttilde}{\widetilde{\bD}}
\safemath{\Bonb}{\bB}
\safemath{\bvec}{\bmb}
\safemath{\Bonbdim}{\ysigdim\times\ysigdim}
\safemath{\noise}{\bmn}
\safemath{\noisedim}{\ysigim}
\safemath{\err}{\bme}
\safemath{\errdim}{\ysigdim}
\safemath{\errset}{\setE}
\safemath{\nerr}{n_e}
\safemath{\delop}{\bP_\errset}
\safemath{\delopc}{\bP_{{\errset}^c}}

\safemath{\cplxi}{\imath}
\safemath{\cplxj}{\jmath}

\safemath{\dict}{\matD}
\safemath{\inputdim}{N}		
\safemath{\outputdim}{M}		
\safemath{\sparsity}{S}	
\safemath{\inputdimA}{{N_a}}	
\safemath{\inputdimB}{{N_b}}	
\safemath{\elemA}{{n_a}}	
\safemath{\elemB}{{n_b}}	
\safemath{\resA}{\matR_a}	
\safemath{\resB}{\matR_b}	
\safemath{\subD}{\matS} 
\safemath{\subA}{\matS_a} 
\safemath{\subB}{\matS_b} 
\safemath{\dicta}{\matA} 	
\safemath{\dictb}{\matB} 	
\safemath{\hollowS}{H}
\safemath{\hollowA}{H_a}
\safemath{\hollowB}{H_b}
\safemath{\cross}{Z}
\safemath{\coh}{\mu_d}			
\safemath{\coha}{\mu_a}			
\safemath{\cohb}{\mu_b}			
\safemath{\mubs}{\nu}	
\safemath{\cohm}{\mu_m} 
\safemath{\dictset}{\setD}	
\safemath{\dictsetp}{\dictset(\coh,\coha,\cohb)}	
\safemath{\dictsetgen}{\dictset_\text{gen}}
\safemath{\dictsetgenp}{\dictsetgen(\coh)}
\safemath{\dictsetonb}{\dictset_\text{onb}}
\safemath{\dictsetonbp}{\dictsetonb(\coh)}

\safemath{\leftside}{U}
\safemath{\rightsideA}{R_a}
\safemath{\rightsideB}{R_b}

\safemath{\indexS}{\setI_S} 

\safemath{\na}{n_a}			
\safemath{\nb}{n_b}			
\safemath{\coeffa}{p_i}	
\safemath{\coeffb}{q_j}	
\safemath{\seta}{\setP}		
\safemath{\setb}{\setQ}     
\safemath{\setw}{\setW}	
\safemath{\setz}{\setZ}	
\safemath{\cola}{\veca}		
\safemath{\colb}{\vecb}		
\safemath{\cold}{\vecd}		
\safemath{\inputvec}{\vecx} 	
\safemath{\error}{\vece}	
\safemath{\noiseout}{\vecz} 	
\safemath{\inputvecel}{x}
\safemath{\inputveca}{\vecx_a}
\safemath{\inputvecb}{\vecx_b}
\safemath{\outputvec}{\vecy}	
\safemath{\lambdamin}{\lambda_{\mathrm{min}}}

\safemath{\elltwo}{\ell_2}
\safemath{\ellone}{\ell_1}
\safemath{\ellzero}{\ell_0}
\safemath{\ellinf}{\ell_\infty}
\safemath{\ellinftilde}{\ell_{\widetilde\infty}}
\safemath{\licard}{Z(\coh,\coha,\cohb)}
\safemath{\xsol}{\hat{x}}
\safemath{\xbord}{x_b}		
\safemath{\xstat}{x_s}		
\safemath{\xstatLone}{\tilde{x}_s}
\safemath{\order}{\mathcal{O}} 
\safemath{\scales}{\Theta} 
\safemath{\ones}{\mathbf{1}} 
\safemath{\zeroes}{\mathbf{0}} 
\safemath{\thlone}{\kappa(\coh,\cohb)} 
\safemath{\constoneA}{\delta} 
\safemath{\constoneB}{\epsilon} 
\safemath{\nlarge}{L}				   
\safemath{\sumlarge}{S_\nlarge}
\safemath{\maxlarger}{P_\nlarge}	   
\safemath{\Pzero}{\textrm{P0}}	
\safemath{\Pone}{\textrm{P1}}
\safemath{\vecfir}{\vecw}			 
\safemath{\vecsec}{\vecz}
\safemath{\elvecfir}{w}              
\safemath{\elvecsec}{z}				 
\safemath{\nlargefir}{n}
\safemath{\normout}{\gamma}
\safemath{\auxfun}{h}
\safemath{\supp}{\textrm{supp}}

\safemath{\indexa}{\ell}
\safemath{\indexb}{r}
\safemath{\indexc}{i}
\safemath{\indexd}{j}

\safemath{\project}{P}

\usepackage{framed}

\renewcommand{\bSigma}{\mathbf{\Sigma}}

\newcommand{\Ie}{\hat{I}}
\renewcommand{\bLambda}{\mathbf{\Lambda}}

\newcommand*\tinygraycircled[1]{\Circled[inner color=white, fill color= gray, outer color=gray]{\footnotesize{\textnormal{#1}}}}

\newcommand{\secret}{\twemoji[height=3mm,trim={0.1mm 0.1mm 0.1mm 0mm}, clip]{game_die}}

\safemath{\sfp}{\textsf{p}}
\safemath{\sfc}{\textsf{c}}

\IEEEoverridecommandlockouts
\allowdisplaybreaks 

\begin{document}
\bstctlcite{IEEEexample:BSTcontrol} 

\title{Jammer-Resilient Time Synchronization\\ in the MIMO Uplink}

\author{
	\IEEEauthorblockN{~~~Gian Marti, Flurin Arquint, and Christoph Studer}
	\thanks{The authors are with the Department of Information Technology and Electrical Engineering, ETH Zurich, Switzerland. (email: marti@iis.ee.ethz.ch, farquin@student.ethz.ch, studer@ethz.ch)}
	\thanks{The work of CS and GM was funded in part by an ETH Grant. 
	Emojis by Twitter, Inc. and other contributors are licensed under CC-BY 4.0}
	\thanks{Simulation code is available on \url{https://github.com/IIP-Group/JASS}.}
}

\maketitle

\begin{abstract}
Spatial filtering based on multiple-input multiple-output (MIMO) processing is a promising approach 
to jammer mitigation. 
Effective MIMO data detectors that mitigate \emph{smart} jammers have recently been proposed, 
but they all assume perfect time synchronization between transmitter(s) and receiver. 
However, to the best of our knowledge, there are no methods for resilient time synchronization 
in the presence of smart jammers. 
To remedy this situation, we propose JASS, the first method that enables reliable time synchronization for the single-user MIMO uplink while mitigating smart jamming attacks.
JASS detects a randomized synchronization sequence based on a novel optimization problem that fits a spatial 
filter to the time-windowed receive signal in order to mitigate the jammer.
We underscore the efficacy of the proposed optimization problem by proving that it ensures successful time synchronization under certain intuitive conditions.
We then derive an efficient algorithm for approximately solving our optimization problem. 
Finally, we use simulations to demonstrate the effectiveness of JASS against a wide range of different jammer~types.
\end{abstract}

\begin{IEEEkeywords}
Jammer mitigation, MIMO, time synchronization.
\end{IEEEkeywords}

\section{Introduction}
\IEEEPARstart{J}{ammers} must be mitigated as they pose a serious threat to the wireless communication infrastructure \cite{threatvectors2021cisa, economist2021satellite, nyt2023invisible}.
Multiple-input multiple-output (MIMO) processing methods enable spatial filtering
and have therefore been recognized as a promising approach for jammer mitigation \cite{leost2012interference, pirayesh2022jamming,  
hoang2021suppression, shen14a, yan2016jamming, marti2023maed, marti2023jmd, do18a, marti2021snips, marti2023universal, zeng2017enabling, jiang2021efficient}.
If the MIMO receiver has more antennas than the jammer (a typical assumption in the context 
of MIMO jammer mitigation), then the jammer interference can only occupy  
a lower-dimensional subspace of the entire signal space, which enables nulling 
by projecting the receive signals on the orthogonal complement of that subspace 
\cite{hoang2021suppression, shen14a, yan2016jamming, do18a, marti2021snips, marti2023maed, marti2023jmd, marti2023universal} 
or by attenuating the signal dimensions that correspond to the jammer's subspace
during equalization \cite{do18a, marti2021snips, marti2023universal, jiang2021efficient, zeng2017enabling}.
These basic mechanisms are surrounded, however, by a myriad of issues, such as
how the relevant quantities of the jammer's spatial signature (i.e., its 
subspace or its spatial covariance matrix) can be estimated at the receiver \cite{pirayesh2022jamming, hoang2021suppression,  
shen14a, yan2016jamming, do18a, marti2023maed, marti2023jmd, marti2021snips, marti2023universal, zeng2017enabling} 
or how a link between the receiver and a legitimate transmitter can be established 
\cite{shen14a, zeng2017enabling, yan2016jamming, shahriar2014phy, zhang2021preamble, lapan2012jamming, 
lichtman2016lte, lichtman20185g, eygi2020countermeasure, el2017lte}.
These latter questions are exacerbated when a strong attack model 
is considered in which the jammer is not assumed to behave in some predictable 
manner (such as steadily transmitting at all times, or when the legitimate transmitter is transmitting)  
that can be exploited for mitigation. Recently, there has been considerable progress on how to estimate the spatial
signature of smart jammers \cite{hoang2021suppression, marti2023maed, marti2023jmd, marti2023universal},
but all of these methods assume that a communication link between the receiver and the legitimate transmitter(s) 
has already been established, i.e., it is typically assumed that (i) the receiver knows
of the existence of the legitimate transmitter(s) and (ii) the receiver 
and transmitter(s) are synchronized in time and frequency.
These assumptions are far from innocent since smart jammers can 
attack also the procedures for link establishment \cite{lapan2012jamming, lichtman2016lte, lichtman20185g}. 
Several methods have been proposed for synchronization in the presence of jamming \cite{la2013protecting, eygi2020countermeasure, el2017lte, 
shen14a, zeng2017enabling, yan2016jamming},
but none of them assume a strong attack model in the sense of \cite{marti2023maed, marti2023jmd, marti2023universal}.

\subsection{Contributions}
We propose JASS (short for Jammer-Aware SynchroniSation), 
the first method that enables reliable time synchronization against arbitrary jamming attacks, including attacks from smart multi-antenna jammers. 
JASS relies on MIMO processing to mitigate jammers via adaptive spatial filtering. 
Specifically, we propose a novel optimization problem for detecting the presence (or absence) of a randomized 
synchronization sequence (that is known to both the transmitter and the receiver, but not to the jammer) at time $\ell$ 
by fitting a spatial filter to the time-windowed receive signal---if the 
optimized objective value exceeds a certain threshold, then the receiver decides that the synchronization sequence
has been transmitted at the corresponding time instant~$\ell$. 
We prove that, under certain intuitive conditions, solving this optimization problem will detect the synchronization 
sequence correctly with probability one. 
Since the optimization problem in question is nonconvex, the JASS synchronization algorithm utilizes 
an efficient procedure (closely related to Dinkelbach's method~\cite{dinkelbach1967nonlinear})
for approximately solving our optimization problem. 
We demonstrate the effectiveness of JASS using simulations that consider a broad range of jammer types.
In particular, JASS is, to the best of our knowledge, the only time synchronization method that can mitigate all of the considered 
jammer types without requiring or exploiting any explicit information about the jammer's type or behavior.

\subsection{Related Work}
The impact of jamming attacks on synchronization has been discussed in 
\cite{shahriar2014phy} and in \cite{lapan2012jamming, zhang2021preamble}.
The latter two works also mention the danger of preamble spoofing to synchronization.
Different protocol specific attacks against LTE and 5G NR, including jamming
and spoofing of synchronization signals, are discussed in \cite{lichtman2016lte, lichtman20185g}. 
As countermeasures against spoofing attacks, reference \cite{zhang2021preamble} 
advocates for randomized preambles. Similarly, 
references \cite{la2013protecting, eygi2020countermeasure} propose to randomize
the in-frame position of synchronization signals, but this countermeasure is 
effective only against jammers that try to be stealthy by \emph{only}
jamming synchronization signals, but not against barrage jammers that jam 
continuously. In \cite{el2017lte}, an adaptive filter with frequency-dependent
weights is proposed as a method for LTE synchronization that is resilient
against partial-band jammers, but this approach remains vulnerable to 
full-spectrum barrage jamming. 
In contrast to these works, our approach mitigates jammers through spatial filtering
and does not depend on any particular jamming strategy (such as jamming \emph{only} synchronization signals 
or partial-band jamming). 
Moreover, since our approach builds upon on a randomized synchronization sequence, it is invulnerable to spoofing. 

The papers \cite{shen14a, zeng2017enabling, yan2016jamming} implement MIMO 
jammer mitigation on Universal Software Radio Peripherals (USRPs) and their implementations include methods for 
time synchronization. 
The work in \cite{shen14a} detects the start of a transmitted frame based on the following insight: If the received signal 
contains both UE and jamming signals, then the receive vectors $\bmy[k]$ and $\bmy[k']$ will in general not be collinear
for $k'\neq k$, while, if the receive signal only contains jamming signals from a single-antenna jammer but no UE signals, 
then $\bmy[k]$ and $\bmy[k']$ will be collinear. 
However, if the jammer is much stronger than the UE, a large number of receive samples is needed to decide between these two hypotheses 
(500 samples are used in \cite{shen14a}).
Moreover, this insight can only be used to mitigate single-antenna jammers, 
since multi-antenna jammers can generate receive vectors
that are not necessarily collinear across different samples. 
The work in \cite{zeng2017enabling} uses a spatial filter to null the jammer interference during the synchronization phase. 
For this, however, one must assume a barrage jammer 
which jams permanently, so that its spatial signature can be estimated from 
the receive signals during a training phase. 
The work in \cite{yan2016jamming} considers a reactive jammer that only starts to jam after the frame start. This work depends on the assumption 
that the jammer's reaction time is sufficiently slow so that the transmit signal 
preamble (which includes synchronization signals) is not jammed. 
In contrast to these works, our method mitigates jammers regardless of whether they have 
one or multiple antennas, and regardless of whether they jam permanently or only at selected instants. 

In summary, we find no synchronization method designed to withstand strong 
jamming attacks in the sense of \cite{marti2023maed, marti2023jmd, marti2023universal}, where the jammer is \emph{not} assumed to behave in some 
predictable way that can be exploited for mitigation. 
Moreover, none of these works on mitigation consider jammers which 
behave in ways that should make it difficult for the receiver to mitigate it
(such as transmitting in erratic bursts, switching between different transmit
antennas to dynamically change its spatial signature, or transmitting only 
at instances when the legitimate transmitter is transmitting).
Our paper fills this gap by (i) assuming a strong attack model in the first 
place and (ii) by specifically considering such adversarial jammers in the 
empirical evaluation of our method.

\subsection{Notation}
Matrices and column vectors are written in boldface uppercase and lowercase letters, respectively.
For a matrix  $\bA\in\opC^{M\times N}$ (and analogously for a vector $\bma\in\opC^M$), 
the conjugate is $\bA^\ast$, the transpose is $\tp{\bA}$, the conjugate transpose is $\herm{\bA}$, 
and the Moore-Penrose pseudo-inverse is $\pinv{\bA}$.
The Frobenius norm is~$\| \cdot \|_F$, and $\|\cdot\|_2$ denotes the spectral norm (for matrices) 
and the Euclidean norm (for vectors), respectively. 
The columnspace and rowspace of $\bA$ are $\textit{col}(\bA)$ and $\textit{row}(\bA)$, respectively.
Horizontal concatenation of two matrices $\bA$ and~$\bB$ is denoted by $[\bA,\bB]$; vertical concatenation is 
 $[\bA;\bB]$.
The $N\times N$ identity matrix is $\bI_N$, $\mathbf{1}_N=[1;\dots;1]\in\opC^N$ is the $N\times1$ all-ones vector, 
and $[N:N']$ denotes the integers from $N$ through $N'$. 
The all-zero vector (whose dimension is implicit) is~$\mathbf{0}$.
The distribution of a circularly-symmetric complex Gaussian vector with covariance matrix $\bC$ is 
$\setC\setN(\mathbf{0},\bC)$.

\section{System Model} \label{sec:system}
\subsection{System Model and Problem Formulation} \label{sec:problem}
We consider the uplink of a multi-antenna system in which a single-antenna UE wants to synchronize to a $B$-antenna~BS 
in presence of an $I$-antenna jammer (or multiple jammers with~$I$ antennas in total) with $I<B$. 
To this end, the UE transmits a \mbox{length-$K$} synchronization sequence $\check\bms=\tp{[\check s_0,\dots,\check s_{K-1}]}\!\in\!\opC^K$
that is known to the BS, but assumed to be unknown to the jammer.
We model the time synchronization problem from the vantage point of the BS, which has to detect the instant at which 
the UE transmits the synchronization sequence. We assume flat fading and 
describe the receive signal at the BS at sample index 
$k\in\opZ_{\geq0}$~as
\begin{align}
	\bmy[k] = \bmh s[k-L]  + \bJ \bmw[k] + \bmn[k]. \label{eq:io}
\end{align}
Here, $\bmy[k]\in \mathbb{C}^{B}$ is the BS receive vector at sample index $k$; $s[\cdot]$ is the UE transmit signal, which we
model as
\begin{align}
	s[k] = \begin{cases}
		0 &: ~~ k<0\\
		\check s_k &:~~ k\in\{0,1,\dots,K-1\}\\
		\text{undefined} &:~~ k\geq K; 
	\end{cases} \label{eq:ue_tx}
\end{align}
$L\in\opZ_{\geq0}$ represents the time at which the UE transmits the synchronization sequence---this is unknown \emph{a priori}
to the BS; $\bmh\in\mathbb{C}^{B}$ and $\bJ\in\mathbb{C}^{B\times I}$ model the channels from the UE to the BS 
and from the jammer to the BS, respectively, where the $i$th column of $\bJ$ holds the channel vector
from the jammer's $i$th antenna to the BS; 
$\bmw[k]\in\opC^I$ is the jammer's transmit signal at sample index $k$; and $\bmn[k]\sim\setC\setN(\mathbf{0},\No\bI_B)$ 
is additive white circularly-symmetric complex Gaussian noise with per-entry variance $\No$, 
which models thermal noise as well as background interference (e.g., inter-cell interference).
Neither $\bmh$ nor $\bJ$ are known at the BS. And, apart from flat fading, we assume no particular channel~model.

The time synchronization problem to be solved is the following: 
Based on a sequence of receive vectors $\{\bmy[k]\}_{k\geq0}$, the BS has to produce an estimate $\hat\ell$ of the 
time index $L$, which we model as a non-negative integer-valued random variable. We consider it a synchronization failure 
if the BS fails to produce the correct estimate $\hat\ell=L$ immediately after\footnote{
That is, the BS must not consider any potential signals that the UE sends \emph{after} transmitting the synchronization sequence
when estimating $L$. For this reason, we do not need to define the UE behavior after the UE transmits the synchronization 
sequence; cf.~\eqref{eq:ue_tx}. One could assume that the UE would transmit signals such as pilots or data symbols 
after the synchronization~sequence.} having received $\{\bmy[k]\}_{0\leq k<K+L}$
and a success otherwise.

\subsection{Assumptions and Limitations} \label{sec:limitations}

We now discuss several assumptions on which our model builds and the limitations that follow from these assumptions. 
In particular, we discuss (i) the fact that the synchronization sequence is sent by the UE and not the BS, 
(ii) the assumptions made on the synchronization sequence~$\check\bms$, and 
(iii) the assumptions made on the jammer's transmit signal~$\bmw[\cdot]$. 

\subsubsection{Uplink synchronization}
In currently deployed cellular systems such as 5G NR, it is the BS (and not the UE)
that transmits synchronization signals~\cite{omri2019synchronization}, so that UEs adapt to the BS's schedule. 
If the UE sends the synchronization signal (as in our model), then the BS has to adapt to the UE's schedule, 
which makes it impossible to coordinate multiple UEs with each other. As a result, serving multiple UEs simultaneously 
becomes much more challenging.
However, mitigating jammers in the downlink is an even more formidable problem than in the uplink, since UEs 
typically have only one or few antennas and thus cannot use spatial filtering for effective jammer mitigation \cite{shen2021beam}.
In this paper, we limit our focus to establishing a robust uplink connection and leave the problem of establishing
a robust downlink connection for future work. 
We note, however, that our setup can straightforwardly be adapted for point-to-point 
MIMO systems in which jammers can be mitigated at both multi-antenna receivers.
We also note that \emph{if} one assumes a scenario in which the UEs are equipped with multiple antennas 
(in particular more antennas than the jammer), then the method of our paper is directly applicable also to downlink synchronization.
In that case, one only has to assign UEs the role of the receiver, and the BS the role of the transmitter.

\subsubsection{Randomized synchronization sequence}
We assume that the synchronization sequence $\check\bms$ is known only to the UE and the BS, but not to the jammer. 
To this end, we draw a random sequence 
$\check\bms\sim\setC\setN(\mathbf{0},\bI_K)$ with unit symbol energy in expectation: $\Ex{}{\|\check\bms\|_2^2}=K$.\footnote{We 
choose a Gaussian sequence to get a probability-one guarantee in \fref{thm:thm}.
If one is willing to accept a non-zero probability of failure, one could also use a random BPSK sequence, for example.}
In practical implementations, both the UE and the BS could construct $\check\bms$ using a Gaussian-like 
pseudo-random function that they seed with a pre-shared secret $\secret$. 
The assumption that the jammer does not know $\check\bms$ is paramount. Otherwise, the jammer could simply transmit
a copy of $\check\bms$---before the UE does---to fool the BS into erroneously detecting the synchronization sequence.
To prevent the jammer from learning the synchronization sequence over time, the synchronization sequence should be different 
after every use. Which synchronization sequence is currently valid could, e.g., be tied to absolute time 
(provided that the UE and the BS have sufficiently accurate local estimates of absolute time). 

Note that updating the synchronization sequence does not need to incur additional 
communication overhead between the UE and the BS. Specifically, a scheme like
the following could be used:
Denote the pre-shared secret by $\secret_0$. 
Then, the BS and the UE generate the initial synchronization sequence $\check\bms_0=f(\secret_0)$ 
with a suitable, non-invertible function $f$,
so that the jammer cannot determine~$\secret_0$ even if it is able to record 
the transmission of $\check\bms_0$.
Then, the BS and the UE use a pseudorandom generator function $g$ (e.g., Xorshift 
\cite{marsaglia2003xorshift}) to generate a new secret
$\secret_1=g(\secret_0)$, and from that the next synchronization sequence
$\check\bms_1=f(\secret_1)$. This process can be repeated endlessly using $\secret_{t+1}=g(\secret_t)$ and
$\check\bms_{t+1}=f(\secret_{t+1})$.

\subsubsection{Strong attack model}
Keeping step with our earlier work \cite{marti2023maed, marti2023jmd, marti2023universal}, 
to enable the mitigation of as many different jammers as possible (and, in particular, smart jammers),
we want to develop a method that relies on \emph{as few assumptions about the jammer's transmit behavior as possible.}
For this reason, we assume neither independence between the processes $\{\bmw[k]\}$ and $\{s[k]\}$, 
nor any particular distribution for $\{\bmw[k]\}$. 
In particular, we do not assume that the jammer behaves in time-invariant manner. 
We only make the following assumption about the jammer's behavior: 
Since the jammer does not know~$\check\bms$, its transmit signal cannot depend non-causally on 
$\check\bms$. More precisely, $\bmw[k+L]$ cannot depend on $\check{s}_{k+1},\dots,\check{s}_{K-1}$.
To work with the strongest possible attack model, we therefore assume only that the jammer's transmit
signal $\bmw[k]$ at time~$k$ cannot depend on $s[k']$ for $k'>k$, but $\bmw[k]$ \emph{can} depend on 
$s[k']$ for all $k'\leq k$. 
In particular, we assume that the jammer can instantaneously recover the UE's transmit signal and transmit any function thereof 
such as, e.g., the synchronization sequence itself. However, the jammer cannot transmit the synchronization 
sequence earlier than the UE. 
Moreover, we do not assume the strength of the jammer interference to be known at the BS. 
We also take into account that the number of jammer antennas $I$ may not be known at the BS. 

\subsubsection{Other limitations} 
Our method is currently limited to frequency-flat communication channels. 
An extension to frequency-selective channels (i.e., channels with inter-symbol interference) 
is left for future work. 
Also, our method currently only enables time synchronization 
but not frequency synchronization (e.g.,  estimating and compensating  carrier frequency and sampling rate offsets). 
This extension is also left for future~work.

\section{Jammer-Resilient Time Synchronization}

\subsection{The Optimization Problem}
We now derive our approach for solving the time synchronization problem outlined in \fref{sec:problem}. 
We start by considering time synchronization for communication systems 
that do \emph{not} suffer from jamming. In that case, a sensible method is 
correlating the receive signal with the synchronization sequence, and declaring the synchronization
sequence to be detected when the correlation exceeds a certain threshold $\tau$. 
Mathematically, this goal can be expressed as solving
\begin{align}
	\argmin_{\ell\in\opZ_{\geq0}} \ell \quad\text{s.t.}\quad
	\left\| \sum_{k=0}^{K-1}\bmy[k+\ell]\check s_k^\ast\right\|_2^2 \geq \tau.
	\label{eq:correlation}
\end{align}
In the presence of jamming, however, the constraint objective $\big\| \sum_{k=0}^{K-1}\bmy[k+\ell]\check s_k^\ast\big\|_2^2$
will be affected heavily by the jammer interference. In particular, the constraint objective increases (with high probability) 
as the interference power increases. In the presence of jamming, solving \eqref{eq:correlation} is 
therefore not reliable for time synchronization. 

A na\"ive strategy to compensate for the constraint objective's dependence on the interference power would be
to normalize the constraint objective with the (windowed) receive energy: 
\begin{align}
	\argmin_{\ell\in\opZ_{\geq0}} \ell \quad\text{s.t.}\quad
	\frac{\left\| \sum_{k=0}^{K-1}\bmy[k+\ell]\check s_k^\ast\right\|_2^2}{\sum_{k=0}^{K-1}\left\|\bmy[k+\ell]\right\|_2^2}
	\geq \tau.
	\label{eq:normalized_correlation}
\end{align}
However, our simulations in \fref{sec:eval} will show that this approach, too, fares poorly in the presence of strong jamming.

In order to achieve jammer-resilient time synchronization, we therefore leverage spatial filtering: 
\emph{If} the jammer channel matrix $\bJ$ were known at the BS, then the BS could project the receive 
signal $\bmy[k]$ onto the orthogonal complement of $\textit{col}(\bJ)$ by multiplying it with the projection matrix $\bP=\bI_B-\bJ\pinv{\bJ}$
to obtain the jammer-free input-output~relation\footnote{
A matrix of the form $\bI_B-\bJ\pinv{\bJ}$, where $\pinv{\bJ}$ denotes the Moore-Penrose inverse of $\bJ$, 
is the orthogonal projection onto the orthogonal complement of $\textit{col}(\bJ)$ \cite{yanai2011projection}. 
If the columns of $\bJ$ are linearly
independent, then $\pinv{\bJ}=\inv{(\herm{\bJ}\bJ})\herm{\bJ}$. However, even if the columns of $\bJ$ are linearly 
dependent, $\pinv{\bJ}$ is well-defined and unique. Thus, $\bP$ is always well-defined.
}
\begin{align}
	\bP\bmy[k] &=  \bP\bmh s[k-L]  + \underbrace{\bP\bJ}_{=\,\mathbf{0}} \bmw[k] + \bP\bmn[k] \\
	&= \bP\bmh s[k-L]  + \bP\bmn[k].
\end{align}
The synchronization sequence could then easily be detected using a correlation-based 
approach analogous to \eqref{eq:correlation} or~\eqref{eq:normalized_correlation}.
In practice, however, $\bJ$ is \emph{not} known at the BS---at least not \emph{a priori}. 
Moreover, we cannot take for granted that the jammer is active during any specific period of time. 
Thus, we cannot estimate~$\bJ$ from any fixed set of receive vectors; 
see also the related discussions in \cite{marti2023maed, marti2023jmd, marti2023universal}.

In order to spatially mitigate the jammer despite not knowing its channel, our approach will be as follows:
Let $\Ie$ be the BS's guess of how many antennas the jammer has at most.
For any possible time index $\ell$ at which the synchronization 
sequence could have been transmitted, we \emph{fit} a projection of the form 
$\tilde\bP=\bI_B - \bA\herm{\bA}, \bA\in\setO^{B\times\Ie}$ 
(where $\setO^{B\times\Ie}$ is the set of~$B\times\Ie$ sized matrices with orthonormal columns, 
the elements of which satisfy $\pinv{\bA}=\herm{\bA}$)
to the windowed receive signal $\bmy[\ell],\dots,\bmy[\ell+K-1]$.
The set of all such projections~$\tilde\bP$ is the Grassmannian manifold\cite{bendokat2024grassmann}
\begin{align}
	\mathscr{G}_{B-\Ie}(\opC^B) = \left\{\bI_B-\bA\herm{\bA}: \bA\in\setO^{B\times\Ie} \right\}\!, \label{eq:def_grassmann}
\end{align}
and the $\tilde\bP\in\mathscr{G}_{B-\Ie}(\opC^B)$ 
with the best fit is the one for which the projected receive sequence $\tilde\bP\bmy[k]$ 
exhibits the highest normalized correlation with the synchronization sequence $\check\bms$.\footnote{
Note that, as long as $\Ie\geq I$, the set $\mathscr{G}_{B-\Ie}(\opC^B)$ always includes a matrix $\tilde\bP$ that satisfies 
$\tilde\bP\bJ=\mathbf{0}$, even if the columns of~$\bJ$ are not orthonormal. 
To see this, let $\bma_1,\dots,\bma_{\textit{rank}(\bJ)}$ be an orthonormal basis of $\textit{col}(\bJ)$, 
let $\bma_{\textit{rank}(\bJ)+1},\dots,\bma_{\Ie}$ be a (possibly empty) set of orthonormal vectors in $\textit{col}(\bJ)^\perp$, 
and let $\bA=[\bma_1,\dots,\bma_{\Ie}]$. Then, it holds that $\bA\in\setO^{B\times\Ie}$, 
so that $\bI_B-\bA\herm{\bA}\in\setO^{B\times\Ie}$, and it also holds that $(\bI_B-\bA\herm{\bA})\bJ=\mathbf{0}$. 
In other words, the restriction in \eqref{eq:def_grassmann} to matrices $\bA$ with \emph{orthonormal} columns does 
not restrict the range of $\mathscr{G}_{B-\Ie}(\opC^B)$ itself. 
}
Mathematically, this fitting of the projection $\tilde\bP$ to the receive signal windowed at time index~$\ell$ can be 
expressed through the following optimization problem: 
\begin{align}
	\max_{\tilde\bP\in\mathscr{G}_{B-\Ie}(\opC^B)} 
    \frac{\left\|\tilde\bP\sum_{k=0}^{K-1}\bmy[k+\ell]\check s_k^\ast\right\|_2^2}
	{\sum_{k=0}^{K-1}\left\|\tilde\bP\bmy[k+\ell]\right\|_2^2}. \label{eq:fit}
\end{align}
By defining the windowed receive matrix 
\begin{align}
    \bY_\ell = [\bmy[\ell],\bmy[\ell+1],\dots,\bmy[\ell+K-1]] \in \opC^{B\times K},
\end{align}
we can rewrite \eqref{eq:fit} as
\begin{align}
    \max_{\tilde\bP\in\mathscr{G}_{B-\Ie}(\opC^B)} 
    \frac{\|\tilde\bP\bY_\ell\check \bms^\ast\|_2^2}{\|\tilde\bP\bY_\ell\|_F^2}.
    \label{eq:matrix_constraint}
\end{align}
Concerning the objective of \eqref{eq:matrix_constraint}, we define 
$\frac{\|\tilde\bP\bY_\ell\check \bms^\ast\|_2^2}{\|\tilde\bP\bY_\ell\|_F^2}\triangleq0$
if $\|\tilde\bP\bY_\ell\check \bms^\ast\|_2^2=0$ and $\|\tilde\bP\bY_\ell\|_F^2=0$.
We propose to estimate the time index $L$ in \eqref{eq:io} by the lowest 
index $\ell$ for which the solution of the optimization problem in \eqref{eq:fit} exceeds some threshold $\tau$:
\begin{mdframed}[style=csstyle,frametitle={The JASS Optimization Problem:}]
\vspace{-0.25cm}
\begin{align}
    \argmin_{\ell\in\opZ_{\geq0}} ~\ell \!\quad \textnormal{s.t.}\quad\!\! 
     \max_{\tilde\bP\in\mathscr{G}_{B-\Ie}(\opC^B)} 
    \frac{\|\tilde\bP\bY_\ell\check \bms^\ast\|_2^2}{\|\tilde\bP\bY_\ell\|_F^2}
    \geq \tau.\!\label{eq:opt_problem}
\end{align}
\vspace{-0.05cm}
\end{mdframed}

Before we provide an algorithm for (approximately) solving the problem in \eqref{eq:opt_problem}, 
we show the basic soundness of our approach. 
In particular, we prove that the solution to \eqref{eq:opt_problem} gives the correct time index $L$
with probability one if the following conditions hold:\footnote{
\fref{thm:thm} is not constructive: It proves that the solution of \fref{eq:opt_problem} 
exhibits a certain property, but it does not show how to compute that solution.
}
\begin{enumerate}
	\item The noise $\bmn[k]$ in \eqref{eq:io} is negligible, i.e., $\No=0$.
	\item The UE channel vector $\bmh$ is not contained in the column space of $\bJ$, i.e., $\bmh\notin\textit{col}(\bJ)$.
	\item The BS's guess $\Ie$ of the number of jammer antennas satisfies $I\leq\Ie<B$.
	\item The synchronization-sequence length satisfies \mbox{$K\!>\!I\!+\!1$.}
	\item The threshold is set to $\tau=\|\check\bms\|_2^2$. 
\end{enumerate}
\begin{thm} \label{thm:thm}
    Under the stated conditions 1--5, with probability one, the optimization problem in \eqref{eq:opt_problem}
    has the solution~$\ell=L$.
\end{thm}

The proof is in \fref{app:proof}.
\fref{thm:thm} reveals a number of desirable features of our approach. 
First, the optimal value of the objective in \eqref{eq:matrix_constraint} does not depend 
on the gain of the UE channel $\bmh$, which may not be known at the BS, but only on the Euclidean 
norm of the synchronization sequence, which is known.
Second, the BS does not need to know the true number~$I$ of jammer antennas. 
All that is required is that the BS's estimate $\Ie$ of $I$ is not an underestimate 
and that it is smaller than the number of BS antennas $B$. 
Third, our approach does not depend on a particular channel model (except for the flat-fading assumption); we only require that the UE channel vector is not contained in the span of the jammer's channel matrix. 
Note that this assumption holds with probability one for all physically realistic channel models that do not contain mass points, and in which the jammer channel is independent of the UE channel (provided that $I<B$).
Note also
that any jammer mitigation method based on spatial filtering depends on such an assumption---otherwise, 
a spatial filter that nulls the jammer interference also nulls the UE signal.
Finally, and most importantly, successful time synchronization does not depend on the jammer's behavior
(including its power). 
We also point out that, while \fref{thm:thm} depends on the absence of noise, 
our simulation results in \fref{sec:eval} (which \emph{do} consider noise) 
demonstrate that the proposed method performs extremely well, even in the presence of significant noise. 

\subsection{The Algorithm}
We now derive an efficient approximate algorithm, called the \emph{JASS algorithm}, for solving the JASS optimization problem 
postulated in \eqref{eq:opt_problem}.
A natural way for solving \eqref{eq:opt_problem} is to solve the ``inner'' problem \eqref{eq:matrix_constraint} for every 
$\ell=0,1,\dots$, and to terminate as soon as the resulting objective value for a given $\ell$ exceeds the threshold $\tau$. 
However, the fractional nature of the constraint 
objective~$\frac{\|\tilde\bP\bY_\ell\check \bms^\ast\|_2^2}{\|\tilde\bP\bY_\ell\|_F^2}$
in \eqref{eq:opt_problem} makes solving this ``inner'' problem~\eqref{eq:matrix_constraint} difficult.\footnote{
The fact that the constraint set $\mathscr{G}_{B-\Ie}(\opC^B)$ is not convex is not the main source of difficulty here, 
because optimization problems over the Grassmannian manifold often have a closed-form solution, see e.g. \cite{marti2023jmd}.
}
The JASS algorithm therefore approximately solves~\eqref{eq:matrix_constraint} for every $\ell=0,1,\dots$ until, for some $\ell$, 
the optimized objective value exceeds $\tau$, and then terminates with $\hat\ell=\ell$.
Note that in the presence of noise, i.e., $\No>0$, 
the threshold $\tau$ should be treated as a tuning parameter 
and must not necessarily be set equal to~$\|\check\bms\|_2^2$. 
Instead, its choice can be determined empirically.

For the derivation of the JASS algorithm, we start by rewriting the optimization problem in \eqref{eq:matrix_constraint} as 
\begin{align}
    \max_{\bA\in\setO^{B\times\Ie}}
    \frac{\|\bmc_\ell\|_2^2 - \|\herm{\bA}\bmc_\ell\|_2^2}{\|\bY_\ell\|_F^2 - \|\herm{\bA}\bY_\ell\|_F^2}, 
    \label{eq:opt_reformulated}
\end{align}
where we use $\bmc_\ell=\bY_\ell\check\bms^\ast$ and $\tilde\bP=\bI_B-\bA\herm{\bA}$, 
so that the optimization space is now parametrized by $\bA\in\setO^{B\times\Ie}$.
Note that both the nominator and the denominator of \eqref{eq:opt_reformulated} are nonnegative. 
Imagine now that we know the optimal objective value of~\eqref{eq:opt_reformulated}, 
which we denote as $\gamma\geq0$. A matrix $\bA\in\setO^{B\times\Ie}$ is a maximizer for 
\eqref{eq:opt_reformulated} if and only if it is a maximizer for 
\begin{align}
	\!\!\max_{\bA\in\setO^{B\times\Ie}} 
	\|\bmc_\ell\|_2^2 - \|\herm{\bA}\bmc_\ell\|_2^2 - \gamma(\|\bY_\ell\|_F^2 - \|\herm{\bA}\bY_\ell\|_F^2),\! 
	\label{eq:dinkelbach}
\end{align}
since the objective in \eqref{eq:dinkelbach} is negative for all $\bA$ for which the objective in \eqref{eq:opt_reformulated}
is smaller than $\gamma$, and zero for all $\bA$ for which the objective in \eqref{eq:opt_reformulated} is equal to $\gamma$. 
Therefore, we will use an \emph{estimate} $\gamma$ of the optimal objective value of \eqref{eq:opt_reformulated}
and optimize the proxy objective \eqref{eq:dinkelbach}, which is easier to solve exactly than~\eqref{eq:opt_reformulated}.
Then, we plug the maximizing matrix~$\bA$ for \eqref{eq:dinkelbach} into the objective of \eqref{eq:opt_reformulated} to obtain 
an estimate of the optimal objective value of \eqref{eq:opt_reformulated}. (Note that this estimate can 
underestimate the optimal objective value of \eqref{eq:opt_reformulated} but cannot overestimate~it.)
This approach is closely related to Dinkelbach's method for nonlinear fractional programs
\cite{dinkelbach1967nonlinear, shen2018fractional}.
However, since the value of~$\gamma$ is typically not known \emph{a priori}, Dinkelbach's method
uses an iterative procedure in which $\gamma$ is updated iteratively according to the objective value from the 
previous iteration \cite{dinkelbach1967nonlinear, shen2018fractional}. 
In our case, we \emph{do know} the optimal objective value, at least in the noiseless case: 
From the proof of \fref{thm:thm}, we know that (in the noiseless case) the optimal value of the objective in 
\eqref{eq:opt_reformulated} is upper-bounded by $\|\check\bms\|_2^2$ and that the upper bound is attained for 
the correct index $\ell=L$, which is the only index for which it matters that we solve the optimization problem 
in \eqref{eq:opt_reformulated} accurately. (It does not matter if we underestimate the optimal objective
value of~\eqref{eq:opt_reformulated} for time indices $\ell\neq L$; cf. \eqref{eq:opt_problem}.)
Therefore, we simply set $\gamma=\|\check\bms\|_2^2$ without iterating as one would do with Dinkelbach's method. 
Our simulation results show that this choice of $\gamma$ is effective, even in the presence of noise; see \fref{sec:eval}.

We now show how to efficiently find the maximizing $\bA$ for the proxy problem in \eqref{eq:dinkelbach}. 
To find the maximizing argument, constant terms can be omitted, so we can solve
\begin{align}
	&\argmax_{\bA\in\setO^{B\times\Ie}}~ \gamma \|\herm{\bA}\bY_\ell\|_F^2 - \|\herm{\bA}\bmc_\ell\|_2^2 \\
	=\,& \argmax_{\bA\in\setO^{B\times\Ie}}~ 
	\text{tr}(\herm{\bA}(\|\check\bms\|_2^2\bY_\ell\herm{\bY_\ell} - \bmc_\ell\herm{\bmc_\ell})\bA), \label{eq:trace}
\end{align}  
where in \eqref{eq:trace}, we have inserted $\gamma=\|\check\bms\|_2^2$ and used properties of the Frobenius norm 
\cite{horn2013matrix}.
Since the matrix \mbox{$\|\check\bms\|_2^2\bY_\ell\herm{\bY_\ell} - \bmc_\ell\herm{\bmc_\ell}$} is symmetric, it can be written as
$\|\check\bms\|_2^2\bY_\ell\herm{\bY_\ell} - \bmc_\ell\herm{\bmc_\ell}=\bM\herm{\bM}$ for some $\bM\in\opC^{B\times B}$, 
and so we can rewrite \eqref{eq:trace} as
\begin{align}
	\argmax_{\bA\in\setO^{B\times\Ie}}~ \|\herm{\bA}\bM\|_F^2. \label{eq:eckhart}
\end{align}
Although non-convex, this problem has a closed-form solution:
The Eckhart-Young-Mirksy theorem \cite{eckart1936approximation, mirsky1960symmetric} implies that \eqref{eq:eckhart} 
is maximized if the columns of $\bA$ consist of the $\Ie$ principal left-singular vectors of $\bM$ 
(i.e., the left-singular vectors that correspond to the $\Ie$ largest singular values),
which are equal to the $\Ie$ principal eigenvectors $\bmq_1,\dots,\bmq_{\Ie}$ of the orthonormal eigendecomposition
\begin{align}
	\|\check\bms\|_2^2\bY_\ell\herm{\bY_\ell} - \bmc_\ell\herm{\bmc_\ell} = \bQ\bLambda\herm{\bQ},
\end{align}
where $\bLambda=\text{diag}(\lambda_1,\dots,\lambda_B)$ with $\lambda_1\geq\dots\geq\lambda_B$.
To summarize, the matrix $\bA\in\opC^{B\times\Ie}$ which maximizes \eqref{eq:dinkelbach} can be
found by computing, either exactly or approximately, the $\Ie$ principal orthonormal eigenvectors of the 
matrix $\|\check\bms\|_2^2\bY_\ell\herm{\bY_\ell} - \bmc_\ell\herm{\bmc_\ell}$. 

For an efficient approximation, we can use a power-iteration method.\footnote{Note that, \label{fnprojection}
despite the subtraction in the second term,
$\|\check\bms\|_2^2\bY_\ell\herm{\bY_\ell} - \bmc_\ell\herm{\bmc_\ell}$ is positive semidefinite.
This matters because a power iteration computes not the eigenvectors corresponding to the largest
eigenvalues, but to the eigenvalues that are \emph{largest in magnitude}. 
If $\|\check\bms\|_2^2\bY_\ell\herm{\bY_\ell} - \bmc_\ell\herm{\bmc_\ell}$ were not positive semidefinite,
then these eigenvalues might be negative, so that a power iteration would not produce the desired result.
However, $\|\check\bms\|_2^2\bY_\ell\herm{\bY_\ell} - \bmc_\ell\herm{\bmc_\ell}$ \emph{is} positive semidefinite, 
which can be seen by writing:\begin{align}
	\|\check\bms\|_2^2\bY_\ell\herm{\bY_\ell} - \bmc_\ell\herm{\bmc_\ell} 
	= \|\check\bms\|_2^2 \bY_\ell \Big(\bI_B - \frac{\check\bms^\ast \tp{\check\bms}}{\|\check\bms\|_2^2}\Big)\herm{\bY_\ell}.
\end{align}
The matrix $\bT\triangleq\bI_K - \frac{\check\bms^\ast \tp{\check\bms}}{\|\check\bms\|_2^2}$ is an orthogonal projection
and therefore satisfies $\bT=\bT\herm{\bT}$ \cite{yanai2011projection}. Hence, 
\begin{align}
	\|\check\bms\|_2^2\bY_\ell\herm{\bY_\ell} - \bmc_\ell\herm{\bmc_\ell} 
	= \|\check\bms\|_2\bY_\ell\bT \herm{(\|\check\bms\|_2\bY_\ell\bT)},
\end{align}
which is a Hermitian positive semidefinite matrix.
}
The resulting time synchronization algorithm, JASS, is summarized in \fref{alg:jass}. 
Note that the left-hand-side of the comparison on line 6 uses the pseudoinverse of $\bA$ 
instead of its conjugate transpose. This is due to the fact that, if a small number of iterations 
$t_\textnormal{max}$ is used in the power iteration method (line 5), then the matrix $\bA$ will 
have columns that are only approximately orthonormal, so $\pinv{\bA}\neq\herm{\bA}$.
Empirically, we observe a significant performance decrease if the conjugate transpose is used instead of 
the pseudoinverse. 

\begin{algorithm}[tp]
  \caption{JASS Algorithm}
  \label{alg:jass}
  \begin{algorithmic}[1]
    \Function{JASS}{$\{\bmy[k]\}_{k\geq0},\check\bms,\tau,\Ie,t_\textnormal{max}$}
    \State $\bY_\ell \leftarrow \big[\bmy[0],\dots,\bmy[K-1]\big]$
    \For{$\ell=0,1,2,\dots$}
        \State $\bmc_\ell = \bY_\ell\check\bms^\ast$
        \State $\bA = \textsc{PrincipEig}\big(\|\check\bms\|_2^2\bY_\ell\herm{\bY_\ell} - \bmc_\ell\herm{\bmc_\ell}, \Ie,t_\textnormal{max}\big)$
        \vspace{0.15mm} 
        \If{$\frac{\big{.\|(\bI_B-\bA\pinv{\bA})\bmc_\ell\|_2^2}}{\big{.\|(\bI_B-\bA\pinv{\bA})\bY_\ell\|_F^2}}\geq\tau$}
        \vspace{0.1mm}
            \State \textbf{return} $\ell$
        \EndIf
        \State $\bY_\ell \leftarrow \big[\bY_\ell[\,:\,,1\!:\!K-1],\bmy[\ell+K]\big]$
    \EndFor
    \EndFunction    
  \end{algorithmic}
\end{algorithm}

\begin{algorithm}[tp]
    \caption{Power iteration method}
    \label{alg:power}
    \begin{algorithmic}[1]
        \Function{PrincipEig}{$\bX,\Ie,t_\textnormal{max}$}
        \State $\bar\bX \leftarrow \bX$
        \For{$i=1,\dots,\Ie$}
        \State \textbf{draw} $\bmq_i\sim\setC\setN(\mathbf{0},\bI_B)$
        \For{$t=1,\dots,t_\textnormal{max}$}
            \State $\bmq_i\leftarrow\bar\bX\bmq_i\big/\|\bar\bX\bmq_i\|_2$
        \EndFor
        \State $\hat\lambda_i=\herm{\bmq_i}\bar\bX\bmq_i$
        \State $\bar\bX \leftarrow \bar\bX - \hat\lambda_i \bmq_i\herm{\bmq_i}$        
        \EndFor
        \State \textbf{return} $[\bmq_1,\dots,\bmq_{\Ie}]$
        \EndFunction
    \end{algorithmic}
\end{algorithm}

\subsection{Interpretability of the Resulting Algorithm}
JASS (\fref{alg:jass}) is the result of our effort to come up with an approximate algorithm for 
the optimization problem in~\eqref{eq:opt_problem}, whose basic sensibility is supported by \fref{thm:thm}. 
With hindsight, however, the JASS algorithm is also interpretable on its own terms: 
According to Footnote \ref{fnprojection}, the matrix $\|\check\bms\|_2^2\bY_\ell\herm{\bY_\ell} - \bmc_\ell\herm{\bmc_\ell}$ 
in line 5 of \fref{alg:jass} can be written as 
$\|\check\bms\|_2^2\bY_\ell\bT \herm{(\bY_\ell\bT)}$, 
where $\bT = \bI_K - \frac{\check\bms^\ast \tp{\check\bms}}{\|\check\bms\|_2^2}$ is the projection onto 
the orthogonal complement of the row (i.e., temporal) subspace spanned by $\tp{\check\bms}$ \cite{yanai2011projection}. 
Since we are only interested in the eigenvectors of $\|\check\bms\|_2^2\bY_\ell\bT \herm{(\bY_\ell\bT)}$, 
the prefactor $\|\check\bms\|_2^2$ is irrelevant and could be omitted. 
\fref{alg:jass} can therefore be understood as follows (see the Reinterpreted Algorithm \ref{alg:reint}, 
which is mathematically equivalent to \fref{alg:jass}).
\begin{itemize}
	\item The receiver first projects (in the temporal domain) the windowed receive matrix $\bY_\ell$ onto the orthogonal
	complement of the row subspace spanned by the synchronization sequence $\tp{\check\bms}$, by computing $\bY_\ell^\orth = \bY_\ell \bT$.
	\item The receiver then estimates (in the spatial domain) the interference subspace by approximating the $\Ie$ dominant eigenvectors of 
	$\bY_\ell^\orth \herm{(\bY_\ell^\orth)}$ (which is equivalent to approximating the $\Ie$ dominant left-singular 
	vectors of $\bY_\ell^\orth$, cf. \cite{marti2023universal}), and collecting them into a matrix $\bA$. 
	The estimated jammer-mitigating projection matrix is therefore $\hat\bP=\bI_B - \bA\pinv{\bA}$. 
	\item Finally, the receiver projects (in the spatial domain) the windowed receive matrix $\bY_\ell$ onto the orthogonal
	complement of the estimated interference subspace by computing $\hat\bP\bY_\ell$ (where $\hat\bP=\bI_B - \bA\pinv{\bA}$), 
	and then tests whether the normalized temporal correlation with the synchronization sequence~$\check\bms$ exceeds 
	the threshold $\tau$.
\end{itemize}
To understand intuitively why this strategy works, consider the following two cases:
\subsubsection*{Case $\ell=L$}
In this case, the projection $\bT$ nulls the UE transmit signal completely. The estimate of the interference subspace $\bA$
will therefore be based only on interference and noise signals. If the jammer interference is \emph{stronger} than the noise, 
then the columns of $\bA$ will essentially form an exact orthonormal basis for the interference subspace. The projection~$\hat\bP$ will therefore perfectly null the jammer interference, so that the normalized correlation of $\hat\bP\bY_\ell$ with 
the synchronization sequence $\check\bms$ will be close to one (with the difference depending only on the noise).
If the jammer interference is \emph{weaker} than the noise (or not present at all), then the synchronization performance will 
be limited by noise rather than jammer interference anyway.
\subsubsection*{Case $\ell\neq L$}
In this case, the synchronization sequence is not present (or not properly aligned), and so the normalized correlation 
between $\hat\bP\bY_\ell$ and $\check\bms$ should not exceed $\tau$. 
Since the jammer does not know $\check\bms$, the correlation between its transmit signal and $\check\bms$ is no better than 
random. Hence, none of the signal parts (UE signal, interference, noise) in $\bY_\ell$ are strongly correlated 
with $\check\bms$. The normalized correlation between~$\hat\bP\bY_\ell$ and $\check\bms$ will therefore (with high probability)
be smaller than $\tau$, regardless of $\hat\bP$.

\makeatletter
\renewcommand*{\ALG@name}{Reinterpreted Algorithm}
\makeatother
\setcounter{algorithm}{0}
\begin{algorithm}[tp]
  \caption{JASS Algorithm}
  \label{alg:reint}
  \begin{algorithmic}[1]
    \Function{JASS}{$\{\bmy[k]\}_{k\geq0},\check\bms,\tau,\Ie,t_\textnormal{max}$}
 	\State $\bT = \bI_K - \check\bms^\ast \herm{(\check\bms^\ast)}/\|\check\bms\|_2^2$
    \State $\bY_\ell \leftarrow \big[\bmy[0],\dots,\bmy[K-1]\big]$
    \For{$\ell=0,1,2,\dots$}
    	\State $\bY_\ell^\orth = \bY_\ell \bT$
        \State $\bA = \textsc{PrincipEig}\big(\bY_\ell^\orth \herm{(\bY_\ell^\orth)}, \Ie,t_\textnormal{max}\big)$
        \State $\hat\bP = \bI_B - \bA\pinv{\bA}$
        \vspace{0.15mm} 
        \If{$\frac{\big{.\|\hat\bP\bY_\ell\check\bms^\ast\|_2^2}}{\big{.\|\hat\bP\bY_\ell\|_F^2}}\geq\tau$}
        \vspace{0.1mm}
            \State \textbf{return} $\ell$
        \EndIf
        \State $\bY_\ell \leftarrow \big[\bY_\ell[\,:\,,1\!:\!K-1],\bmy[\ell+K]\big]$
    \EndFor
    \EndFunction    
  \end{algorithmic}
\end{algorithm}

\section{Experimental Evaluation} \label{sec:eval}

\subsection{Simulation Setup} \label{sec:setup}

In order to evaluate JASS, we simulate a single-user MIMO system in the presence of a multi-antenna jammer. 
Unless noted otherwise, the system parameters are as follows: 
The number of BS receive antennas is $B=16$, the number of jammer antennas is $I=4$, the BS's guess of the number
of jammer antennas is $\Ie=4$, the length of the synchronization sequence is $K=16$, and the number of 
power iterations used by JASS (and of baseline methods) for eigenvector approximations is $t_\textnormal{max}=4$. 
We model the distribution of the time index~$L$ at which the UE transmits the synchronization sequence 
as a geometric distribution with parameter $p=\frac{1}{K^2}$ (so that $\Ex{}{L}=K^2-1$).
A geometric distribution for $L$ corresponds to the case where, for every index $\ell$, 
if the synchronization sequence has not arrived until index $\ell-1$, then it arrives during the $\ell$th index 
with constant probability $p$.
Our results will be shown in the form of receiver operating characteristic (ROC) curves over the choice
of the threshold parameter $\tau$, which is therefore not a fixed quantity in our experiments. 
For simplicity, we use an i.i.d. Rayleigh fading channel model for which all entries of the UE channel vector~$\bmh$
and the jammer channel matrix~$\bJ$ are i.i.d. complex standard normal $\setC\setN(0,1)$ distributed; to demonstrate the robustness of our approach, different channel models are considered in \fref{sec:channels}.
We emphasize, however, that JASS depends in no way on the assumption of Rayleigh fading channels.
We express the noise power $\No$ via the average receive signal-to-noise ratio (SNR), which we define as
\begin{align}
	\text{SNR} =\frac{\frac{1}{K}\mathbb{E}\big[\|\bmh \tp{\check \bms}\|_F^2]}{\mathbb{E}\big[\|\bmn[k]\|_2^2]} = \frac{1}{N_0},
	\label{eq:snr}
\end{align}
and we characterize the jammer's transmit power by the parameter $\rho$ (see \fref{sec:jammers} for the details). 
Since the expected symbol energy at which the UE transmits the synchronization sequence is equal to one, 
$\rho$ can also be interpreted as the jammer-to-signal-power ratio. 

When evaluating time synchronization algorithms, we distinguish between two types of errors: 
false positives ($\hat\ell<L$), where the synchronization algorithm declares the synchronization 
sequence to be detected before it has been transmitted, and false negatives, where the synchronization algorithm
does not produce the correct index estimate $\hat\ell=L$ after having observed $\{\bmy[k]\}_{0\leq k<K+L}$.
The threshold parameter $\tau$ (cf. \eqref{eq:opt_problem}) introduces a tradeoff between the 
probability of a false positive and a false negative decision. 
To evaluate the performance of JASS in comparison with two baselines (described in \fref{sec:baselines}), 
we use a Monte--Carlo simulation for estimating the ROC curve between the false positive rate (FPR) 
and the false negative rate (FNR) when varying the threshold parameter~$\tau$. 
We also compute the total error rate (TER), which is nothing but the rate of combined negative and positive errors, 
defined as $\text{TER}=\text{FPR}+\text{FNR}$.
The distribution of synchronization errors $L-\hat\ell$ is investigated
in \fref{app:small}.

\subsection{Jammer Types} \label{sec:jammers}

To highlight the universality of JASS against all types of jammers, we consider six different
jammer types. We emphasize that JASS is given no information about which 
type of jammer it is facing and that it executes exactly the same algorithm 
(outlined in \fref{alg:jass}) against all of them.
The considered jammer types are as follows:
\subsubsection*{\tinygraycircled{1} Barrage jammer}
This jammer transmits i.i.d. random vectors $\bmw[k]\sim\setC\setN(\mathbf{0},\rho\bI_I)$ for all time indexes $k$. 
\subsubsection*{\tinygraycircled{2} Reactive jammer}
This jammer transmits i.i.d. random vectors $\bmw[k]\sim\setC\setN(\mathbf{0},\rho\bI_I)$ during those time indexes
$k$ where the UE transmits the synchronization sequence (i.e., for $k\in[L:L+K-1]$) and is silent during 
all other time indexes.
\subsubsection*{\tinygraycircled{3} Spoofing jammer}
This jammer transmits an amplified copy of the synchronization sequence $\check\bms$ on all antennas as soon as the UE transmits
the synchronization sequence (i.e., it retransmits the synchronization sequence with zero delay): 
\begin{align}
	\bmw[k] = \begin{cases}
		\sqrt{\rho}\check{s}_{k-L}\mathbf{1}_I &:~~ k\in[L:L+K-1]\\
		\mathbf{0} &:~~ \text{otherwise}.
	\end{cases}
\end{align}
\subsubsection*{\tinygraycircled{4} Delayed spoofing jammer}
This jammer retransmits an amplified copy of the synchronization sequence $\check\bms$ on all antennas with one sample 
delay compared to the UE:
\begin{align}
	\bmw[k] = \begin{cases}
		\sqrt{\rho}\check{s}_{k-L-1}\mathbf{1}_I &:~~ k\in[L+1:L+K]\\
		\mathbf{0} &:~~ \text{otherwise}.
	\end{cases}
\end{align}
\subsubsection*{\tinygraycircled{5} Erratic jammer}
This jammer alternates between jamming for bursts whose lengths are independently and uniformly distributed from $1$ to $K$, 
and being silent for periods whose lengths are also independently and uniformly distributed from \mbox{$1$ to $K$.}
During its active jamming phases, the jammer transmits i.i.d. random vectors 
$\bmw[k]\sim\setC\setN(\mathbf{0},\rho\bI_I)$.
\subsubsection*{\tinygraycircled{6} Antenna-switching jammer}
For periods whose lengths are independently and uniformly distributed from $1$ to $K$, 
the antenna-switching jammer selects a random 
subset $\setI\subset[1:I]$ of its antennas and then transmits i.i.d. random vectors 
that satisfy \mbox{$\bmw_{\setI}[k]\sim\setC\setN(\mathbf{0},\rho\bI_{|\setI|})$}
(where $\bmw_{\setI}$ denotes the elements of $\bmw$ that are indexed by $\setI$) and
$\bmw_{[1:I]\setminus\setI}[k]=\mathbf{0}$ for the duration of that period, before 
selecting an independent new subset of antennas in the next period.

\subsection{Baseline Methods} \label{sec:baselines}
We compare JASS in~\fref{alg:jass} with two baseline methods for time synchronization:
\subsubsection*{Unmitigated}
This baseline estimates $\hat\ell$ according to \eqref{eq:normalized_correlation}. 
\subsubsection*{Baseline JASS (BAJASS)}
BAJASS tries to mitigate the jammer as follows:
Before computing the normalized correlation as in \eqref{eq:normalized_correlation}, it
nulls the $\Ie$ strongest spatial dimensions of the receive signal windowed to 
\mbox{$\bY_\ell = [\bmy[\ell],\bmy[\ell+1],\dots,\bmy[\ell+K-1]]$,}
which it attributes to jammer interference. That is, BAJASS estimates $\hat\ell$ according to 
\begin{align}
	\argmin_{\ell\in\opZ_{\geq0}} \ell \quad\text{s.t.}\quad
    \frac{\|(\bI_B-\bA_\ell\pinv{\bA_\ell})\bY_\ell\check \bms^\ast\|_2^2}{\|(\bI_B-\bA_\ell\pinv{\bA_\ell})\bY_\ell\|_F^2}
	\geq \tau,
\end{align}
where the columns of $\bA_\ell\in\opC^{B\times\Ie}$ consist of approximations of the $\Ie$ principal 
left-singular vectors of the windowed receive matrix $\bY_\ell$, which can be approximated using a 
power iteration method similar to \fref{alg:power}. The estimated number $\Ie$ of jammer antennas 
and the number of power iterations $t_\textnormal{max}$ are the same as for the JASS algorithm.
We note that the complexity of this algorithm is comparable to JASS. 

\subsection{Results}

\subsubsection{Performance against barrage jammers} \label{sec:barrage}
\begin{figure*}[tp]
    \centering
    \subfigure[$\rho=0$\,dB barrage jammer]{
        \includegraphics[width=0.23\textwidth]{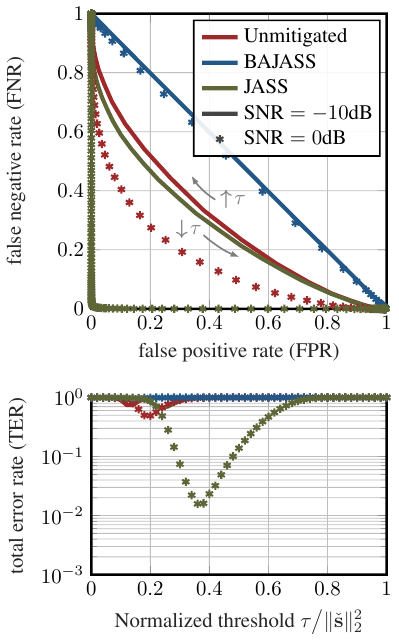}
        \label{fig:0dbbarrage}
    }
    \hspace{15mm}
    \subfigure[$\rho=10$\,dB barrage jammer]{
        \includegraphics[width=0.23\textwidth]{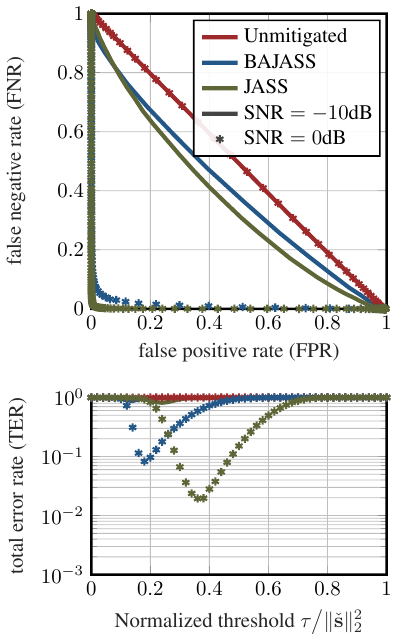}
    }
    \hspace{15mm}
    \subfigure[$\rho=30$\,dB barrage jammer]{
        \includegraphics[width=0.23\textwidth]{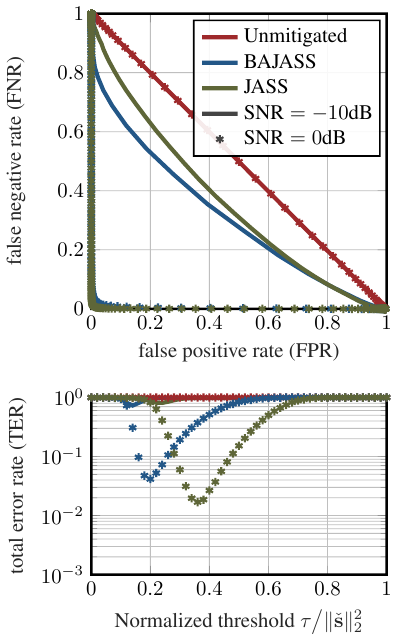}
    }
    \caption{Performance against barrage jammers (\tinygraycircled{1}) with different transmit powers.
    The jammers have $I=4$ antennas; the receiver assumes $\Ie=4$.}
    \label{fig:barrage}
\end{figure*}
\begin{figure*}[tp]
    \centering
    \subfigure[Reactive jammer with $\rho=0$\,dB]{
        \includegraphics[width=0.23\textwidth]{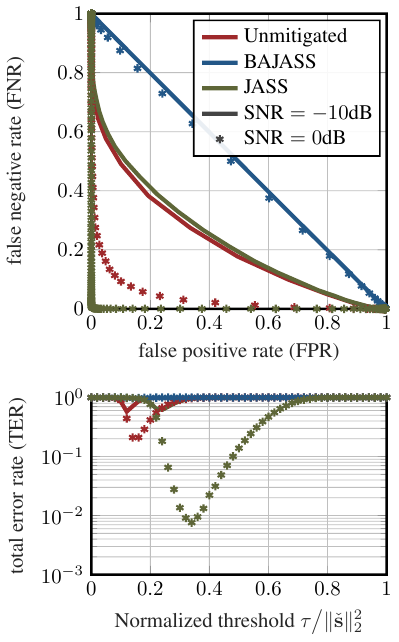}
    }
    \hspace{15mm}
    \subfigure[Reactive jammer with $\rho=10$\,dB]{
        \includegraphics[width=0.23\textwidth]{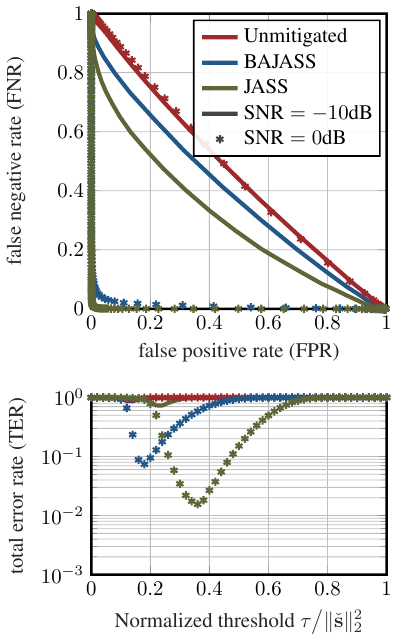}
    }
    \hspace{15mm}
    \subfigure[Reactive jammer with $\rho=30$\,dB]{
        \includegraphics[width=0.23\textwidth]{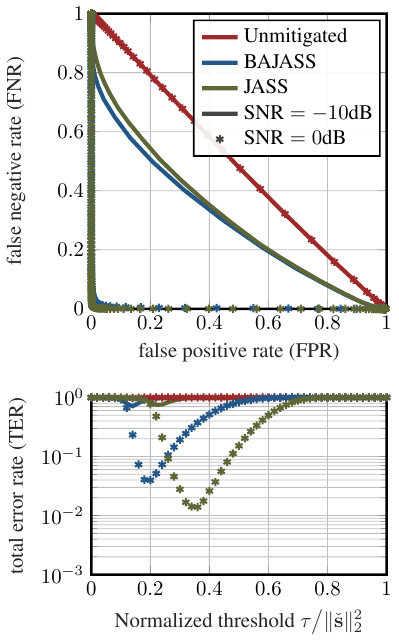}
    }
    \caption{Performance against reactive jammers (\tinygraycircled{2}) with different transmit powers.
     The jammers have $I=4$ antennas; the receiver assumes $\Ie=4$.}
     \label{fig:onsync}
     \vspace{-2mm}
\end{figure*}
In the first experiment, we evaluate the performance of the JASS algorithm and baselines against barrage jammers with $I=4$ 
antennas (JASS and BAJASS assume $\Ie=4$) and different transmit powers, at two different noise levels 
($\text{SNR}=-10$\,dB and $\text{SNR}=0$\,dB). The results are shown in \fref{fig:barrage}: 

At an SNR of $0$\,dB, JASS achieves reliable time synchronization regardless of the jammer transmit power: for a sensible 
choice of $\tau=0.375\|\check\bms\|_2^2$, the TER is around $1\%$ for all jammer powers. 
In contrast, the unmitigated baseline has a TER of more than $40\%$ for all $\tau$ even for the weakest 
jammer and a TER close to $100\%$ for the stronger jammers. 
The BAJASS baseline identifies the jammer interference with the strongest spatial dimensions of the receive signal, 
which is a good approximation for strong jammers, but a bad one for weak jammers. 
In fact, if the jammer's interference is weak compared to the UE signal, BAJASS will indadvertently null 
a significant part of the UE receive signal. Thus, we see that BAJASS performs well (but not as good as JASS)
against the stronger jammers, but extremely poorly against weaker~jammers. 

Note that the choice of a good threshold $\tau$ depends on the synchronization method: 
For unmitigated synchronization and BAJASS, a good threshold is $\tau\approx0.2\|\check{\bms}\|_2^2$; 
for JASS, a good threshold is $\tau\approx0.375\|\check{\bms}\|_2^2$.
We also observe that, for the given system dimensions, an SNR of $-10$\,dB is too 
low for reliable synchronization, with the TER remaining significantly above~$50\%$ in all cases and for all considered methods.

\subsubsection{Performance against reactive jammers}
\begin{figure*}[h]
    \centering
    \subfigure[Spoofing jammer \tinygraycircled{3}]{
        \includegraphics[width=0.23\textwidth]{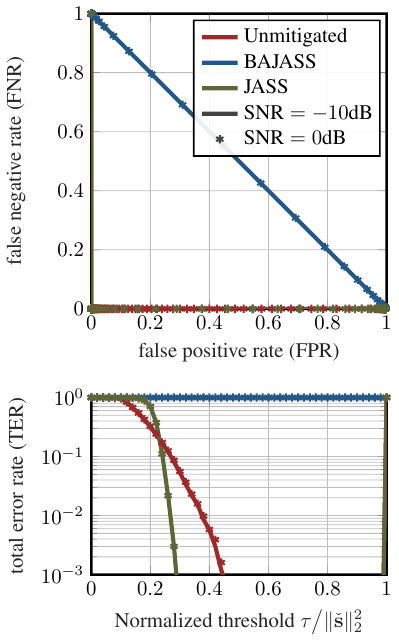}
        \label{fig:spoofing}
    }
    \subfigure[Delayed spoofing jammer \tinygraycircled{4}]{
        \includegraphics[width=0.23\textwidth]{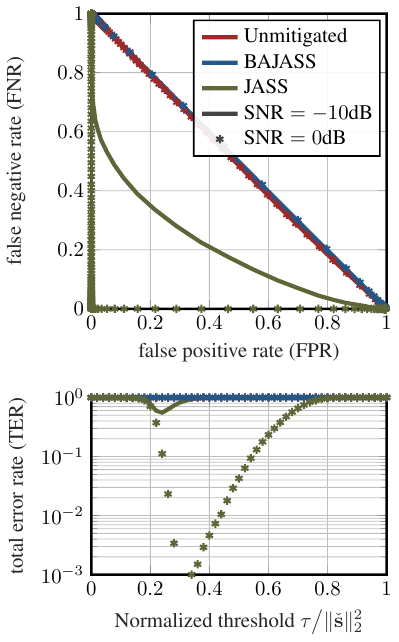}
        \label{fig:delayed}
    }
    \subfigure[Erratic jammer \tinygraycircled{5}]{
        \includegraphics[width=0.23\textwidth]{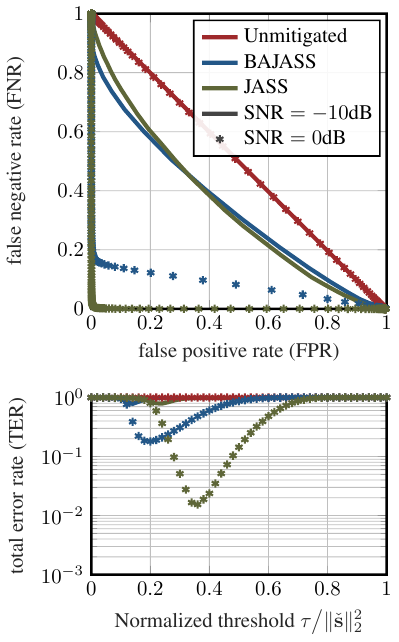}
        \label{fig:erratic}
    }
    \subfigure[Antenna-switching jammer \tinygraycircled{6}]{
        \includegraphics[width=0.23\textwidth]{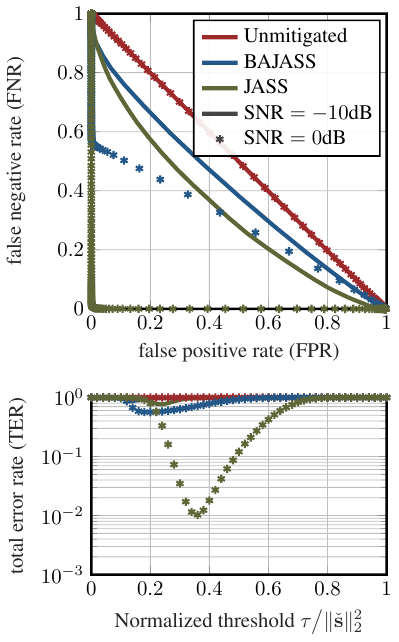}
        \label{fig:dynamic}
    }
    \caption{Performance against different jammers with a transmit power of $\rho=30$\,dB.
    All jammers have $I=4$ antennas; the receiver assumes $\Ie=4$.}
    \label{fig:different}
\end{figure*}
In the second experiment, we evaluate the performance against reactive jammers. 
We again assume jammers with $I=4$ antennas (JASS and BAJASS assume $\Ie=4$) and different transmit powers, 
at two different noise levels ($\text{SNR}=-10$\,dB and $\text{SNR}=0$\,dB).
The results are shown in \fref{fig:onsync}:

Overall, the performance is similar to the performance against the barrage jammers from the previous experiment. 
At an SNR of $-10$\,dB, we observe again that reliable synchronization is impossible
with all the considered methods---the noise level is simply too high. 
At an SNR of $0$\,dB, unmitigated synchronization performs reasonably well against the weakest jammer, 
but fails against the stronger jammers. 
BAJASS performs well (but not as good as JASS) against the stronger 
jammers, but fails against weak jammers.
JASS consistently achieves the best synchronization performance, with a TER of around $1\%$
for all jammer powers when the threshold is set to $\tau=0.375\|\check{\bms}\|_2^2$.

\subsubsection{Performance against the other jammers types} \label{sec:different}

\begin{figure*}[tp]
    \centering
    \subfigure[$I=5$, $\Ie=4$]{
        \includegraphics[width=0.23\textwidth]{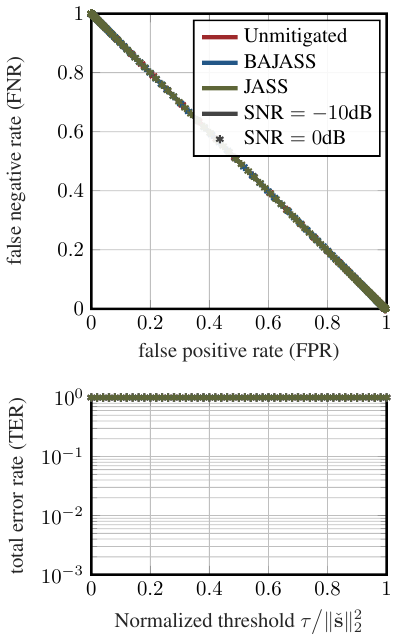}
       	\label{fig:underestimate}
    }       
    \subfigure[$I=3$, $\Ie=4$]{
        \includegraphics[width=0.23\textwidth]{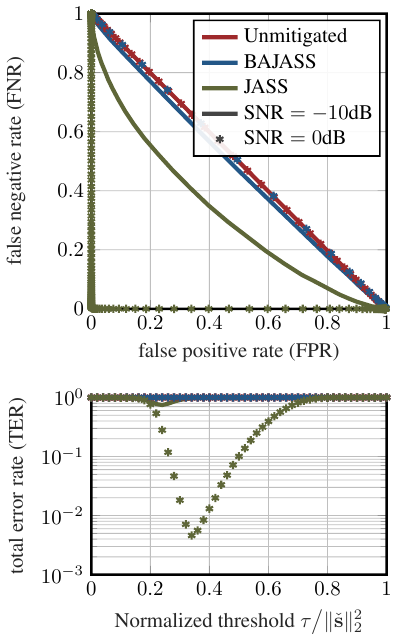}
        \label{fig:overestimate_1}
    }
    \subfigure[$I=1$, $\Ie=4$]{
        \includegraphics[width=0.23\textwidth]{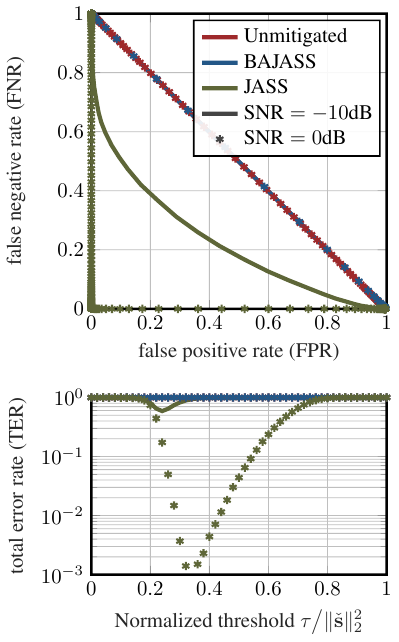}
        \label{fig:overestimate_2}
    }    
    \subfigure[$I=0$, $\Ie=4$ (no jammer)]{
        \includegraphics[width=0.23\textwidth]{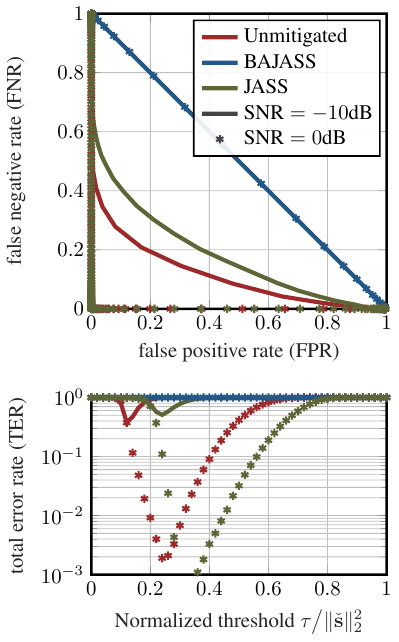}
        \label{fig:no_jammer}
    } 
    \caption{Performance against barrage jammers (\tinygraycircled{1}) with a transmit power of $\rho=30$\,dB
    when there is a mismatch between the number of jammer antennas $I$; the receiver's guess $\Ie$ 
    about the number of jammer antennas.}
    \label{fig:mismatch}
\end{figure*}
We now consider the performance against the other four jammer types from \fref{sec:jammers}. 
In each case, the jammer has $I=4$ antennas and a transmit power of $\rho=30$\,dB (JASS and BAJASS assume $\Ie=4$).
The results are shown in \fref{fig:different}:

\fref{fig:spoofing} shows that both the unmitigated method and JASS achieve virtually errorless performance
against the spoofing jammer, even at $\text{SNR}=-10$\,dB. 
The reason for this increase in synchronization reliability (compared to the barrage or reactive jammers) 
is that the spoofing jammer transmits 
the synchronization sequence \emph{simultaneously} with the UE, 
and so can be seen as helping the receiver detect the synchronization sequence at the correct time index $L$. 
BAJASS, however, fails against the spoofing jammer. The reason for BAJASS's breakdown is as follows: 
for the spoofing jammer, 
when the synchronization sequence is being transmitted ($k\in[L:L+K-1]$), the receive signal can be written as
\begin{align}
	\bY_L &= \bmh\tp{\check{\bms}} + \sqrt{\rho}\bJ\mathbf{1}_I\tp{\check{\bms}} + \bN_L \\
	&= (\bmh+\sqrt{\rho}\bJ\mathbf{1}_I)\tp{\check{\bms}} + \bN_L,
\end{align}
where $\bN_L=[\bmn[L],\dots,\bmn[L+K-1]]$, and 
where $\bmh+\sqrt{\rho}\bJ\mathbf{1}_I$ is a $B\times1$ matrix. Thus, the receive 
signal consists only of a rank-one component in the spatial direction of $\bmh+\sqrt{\rho}\bJ\mathbf{1}_I\in\opC^{B}$, 
plus additive white Gaussian noise. Since BAJASS nulls the $\Ie=4$ strongest spatial dimensions of $\bY_L$ before 
measuring the correlation with the synchronization sequence, it will null the 
spatial subspace $\textit{span}\{\bmh+\sqrt{\rho}\bJ\mathbf{1}_I\}$ (which dominates the noise subspace due to the 
strong component from the jammer), as well as three random dimensions of the noise subspace. 
Since $\textit{span}\{\bmh+\sqrt{\rho}\bJ\mathbf{1}_I\}$ is the spatial subspace that contains synchronization sequence, 
its nulling makes it impossible to detect the synchronizaton sequence afterwards.  

\fref{fig:delayed} shows that JASS is the only method that achieves good performance against the delayed spoofing jammer. 
The unmitigated method fails because the delayed synchronization sequence from the jammer
hinders rather than supports the detection of the synchronization sequence at the correct time index $L$. 
BAJASS fails because the receive signal during the synchronization sequence can now be written as
\begin{align}
	\bY_L &= \bmh\tp{\check{\bms}} + \sqrt{\rho}\bJ\mathbf{1}_I [0,\check{s}_0, \dots, \check{s}_{K-2}]
	 + \bN_L, 
\end{align}
where $\sqrt{\rho}\bJ\mathbf{1}_I$ is a $B\times1$ matrix.
In contrast to the scenario of a spoofing jammer, the jammer does not occupy the same spatial subspace as the UE 
signal anymore (due to the delay in the spoofed synchronization sequence), 
but the jammer itself still occupies only a one-dimensional subspace of the 
receive signal. BAJASS will thus null $\textit{span}\{\sqrt{\rho}\bJ\mathbf{1}_I\}$, 
as well as the three next strongest dimensions of the receive signal. Since there are no 
other dimensions that contain jammer interference, the only other dimension that stands out from 
the spatially uniform noise is the spatial subspace of the UE signal, $\textit{span}\{\bmh\}$, 
which will therefore also be nulled (along with two more random dimensions of the noise subspace).

\fref{fig:erratic} and \fref{fig:dynamic} show the performance against the erratic jammer and
the antenna-switching jammer, respectively.  
The unmitigated method fails against both jammers. 
BAJASS performs somewhat better than against the spoofing jammers, but still exhibits a TER of more 
than $10\%$ (or more than $50\%$, respectively) regardless of the threshold $\tau$. 
JASS performs well also against these two jammers, with a TER of around $1\%$ for a threshold of
$\tau=0.375\|\check{\bms}\|_2^2$ (at an SNR of $0$\,dB).

\subsubsection{Performance under jammer antenna mismatch ($\Ie\neq I$)} \label{sec:mismatch}
So far, in all of the previous experiments, JASS and BAJASS have operated using a correct guess $\Ie=I=4$ of the 
jammer's number of transmit antennas. However, assuming that the receiver knows the exact number 
of jammer antennas is likely not realistic in practice. 
Our theoretical result (\fref{thm:thm}) implies  that JASS should work even when $\Ie>I$.
In this experiment, we therefore evaluate cases where there is a mismatch between the receiver's guess $\Ie$ 
for the number of jammer antennas and the true number of jammer antennas~$I$. 
The results are shown in \fref{fig:mismatch}. All jammers are barrage jammers with a transmit power of $\rho=30$\,dB.

As expected from \fref{thm:thm}, JASS performs well when the number of jammer antennas is overestimated 
(see \fref{fig:overestimate_1} and \fref{fig:overestimate_2}).\footnote{
Overestimating the number of jammer antennas \emph{does} come at a price.
An estimated number of $\Ie$ jammer antennas means that an $\Ie$-dimensional subspace is nulled at the receiver. 
In uncorrelated Rayleigh fading, the effect of nulling an $\Ie$-dimensional subspace
on the legitimate signal is mathematically equivalent to the loss of $\Ie$ receive antennas; see, e.g.,~\cite[Prop.\ 4]{marti2024fundamental}. Thus, overestimating the number of jammer antennas 
leads to decreased signal power and diversity.
}
In contrast, BAJASS fails and achieves a TER close to $100\%$ because it inadvertantly uses the surplus of assumed jammer 
dimensions to null the UE signal (analogous to the case of the delayed spoofing jammer; cf. \fref{sec:different}).

In contrast, if the number of jammer antennas is underestimated (\fref{fig:underestimate}), \emph{all} methods fail. 
This failure comes as no surprise: if the number of jammer antennas exceeds the receiver's estimate, and if the jammer 
transmits independent interference on all its antennas, then the dimension of the 
spatial interference subspace exceeds the number of dimensions that are nulled by JASS and BAJASS.
As a result, the interference cannot be removed completely by the subspace nulling of JASS and BAJASS.
Since a significant part of the jammer interference remains, all methods exhibit a TER of very close to $100\%$, 
regardless of the threshold $\tau$. 

The final case of interest is the case where \emph{no} jammer is interfering (\fref{fig:no_jammer}), 
which can be viewed as a special case of overestimating the number of jammer antennas. 
Unsurprisingly, the unmitigated method performs well in this scenario. 
Even in the jammerless scenario, however, the performance of JASS is comparable to that of the unmitigated performance, 
showing that JASS can also be used in scenarios where it is uncertain if a jamming attack is (or will be) occurring or not. 
In contrast, BAJASS fails again because it nulls the UE signal 
(analogous to the case of the delayed spoofing jammer; cf. \fref{sec:different}).

\subsubsection{Results summary}
\begin{figure*}[tp]
    \centering
    \subfigure[synchronization sequence length~$K$\!\!]{
        \includegraphics[width=0.23\textwidth]{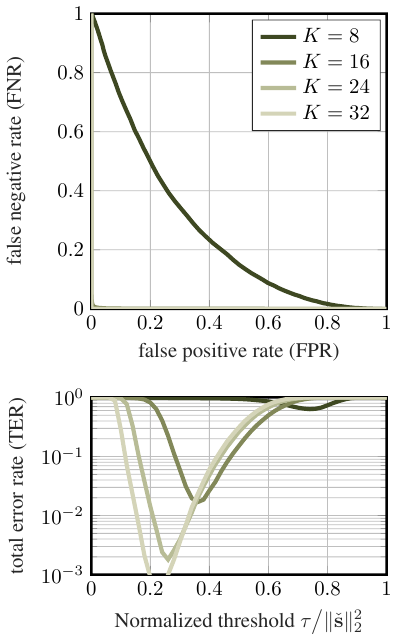}
        \label{fig:vary_K}
    }
    \hspace{15mm}
    \subfigure[receive antennas $B$]{
        \includegraphics[width=0.23\textwidth]{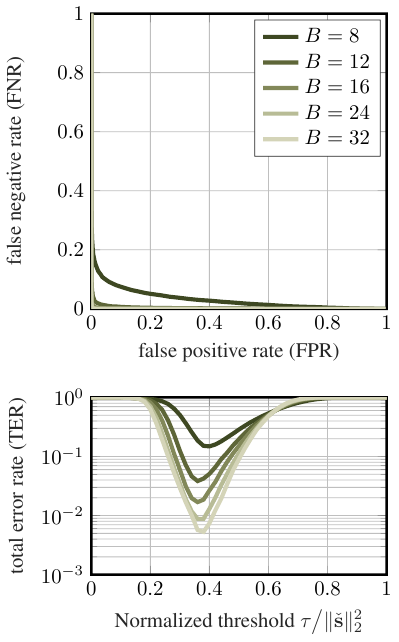}
        \label{fig:vary_B}
    }
    \hspace{15mm}
    \subfigure[noise power $\No$]{
        \includegraphics[width=0.23\textwidth]{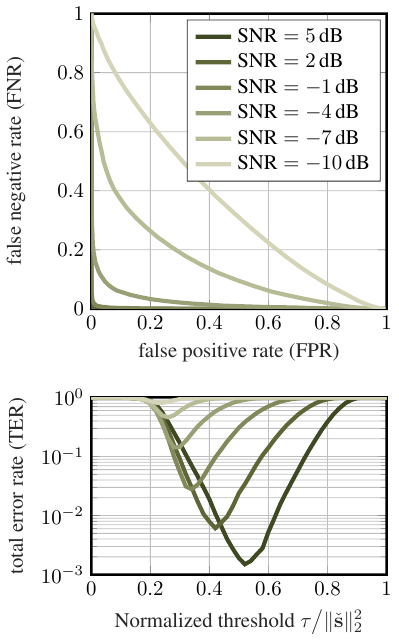}
        \label{fig:vary_SNR}
    }
    \subfigure[jammer power $\rho$]{
        \includegraphics[width=0.23\textwidth]{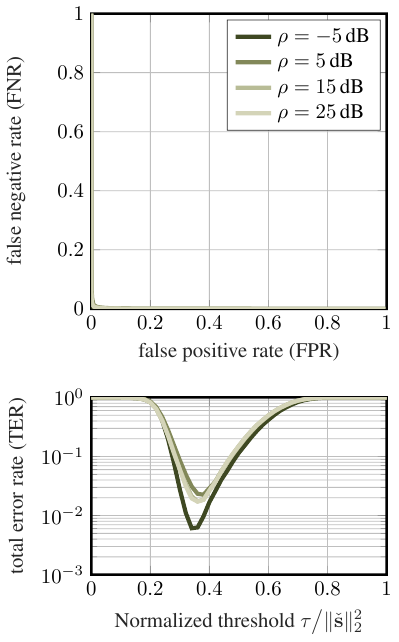}
        \label{fig:vary_rho}
    }
    \hspace{15mm}
    \subfigure[(estimated) jammer antennas \mbox{$I=\hat{I}$}\!\!\!]{
        \includegraphics[width=0.23\textwidth]{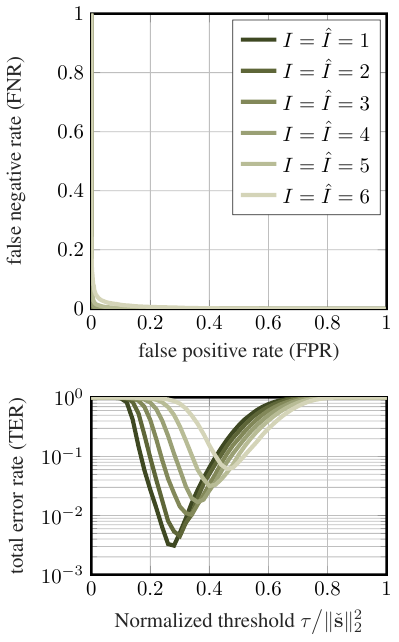}
        \label{fig:vary_I}
    }    
    \hspace{15mm}
    \subfigure[power iterations $t_\textnormal{max}$]{
        \includegraphics[width=0.23\textwidth]{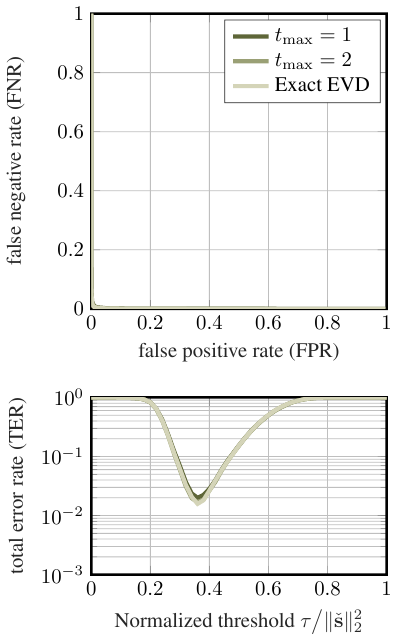}
        \label{fig:vary_tmax}
    }    
    \caption{Ablation studies. Default values are $K=16$, $B=16$, $\textit{SNR}=0\,\text{dB}$, $\rho=30\,\text{dB}$, $I=\hat{I}=4$, $t_\textnormal{max}=4$. The jammer is a barrage jammer in all cases.}
    \label{fig:ablation}
\end{figure*}
Our results show that JASS works reliably against all of the considered types of jammers and, at an SNR of $0$\,dB achieves a TER of around $1\%$ or less against all jammers (as long as 
the number of jammer antennas $I$ satisfies $I\leq\Ie$). In contrast, the unmitigated method
fails against all types of strong jammers except for the spoofing jammer; and BAJASS fails 
against weak jammers, spoofing jammers, delayed spoofing jammers, erratic jammers, antenna-switching jammers, 
as well as when the exact number of jammer antennas~$I$ is not known,~$I\neq\Ie$. 

We conclude this evaluation by pointing out an advantageous characteristic of JASS. The choice of a good threshold does 
not depend on the type of jammer; $\tau=0.375\|\check{\bms}\|_2^2$ seems to be a universally good choice for all of the considered jammers. 
This independence of the threshold to the jammer type is important because, in a real-world scenario, the behavior of an attacking 
jammer is typically not known in~advance.

\subsection{Performance of JASS Under Varying System Parameters}

After having demonstrated the effectiveness of JASS against different types of jamming attacks, 
we analyze its performance under varying system parameters.
The results are shown in \fref{fig:ablation}.
In each experiment, one system parameter is varied. 
The system parameters that are held constant for a given experiment are set to the following values: 
The synchronization sequence length is $K=16$, the number of BS antennas is $B=16$, 
the noise power $\No$ corresponds to $\text{SNR}=0$\,dB, the jammer is a barrage jammer, 
the jammer transmit power is $\rho=30$\,dB, the (estimated) number of jammer antennas is $\Ie=I=4$, 
and the number of power iterations is $t_{\textrm{max}}=4$.
In the first experiment, we vary the synchronization sequence length $K$ (see \fref{fig:vary_K}).
Unsurprisingly, the synchronization performance improves with~$K$.\footnote{Note
that the expected arrival time of the synchronization sequence increases quadratically with $K$ 
in our experiments; cf. \fref{sec:setup}.}
Note that the value of the optimal threshold $\tau$ changes with $K$ as well. This
can easily be taken into account at design time of a system, since the receiver knows $K$. 

In the second experiment, we vary the number of receive antennas $B$ (see \fref{fig:vary_B}). 
Again, unsurprisingly, the performance improves with $B$. In this case, $B$ appears 
to have no influence on the choice of the optimal threshold parameter $\tau$. 

In the third experiment, we vary the noise power $\No$ by varying the SNR (see \fref{fig:vary_SNR}). 
The relation between SNR and~$\No$ is described by \eqref{eq:snr}.
Naturally, the performance improves as the noise power decreases. Moreover, 
with decreasing noise power, the optimal threshold parameter moves closer to the optimal objective value $\|\check{\bms}\|_2^2$
from \fref{thm:thm}. 

In the fourth experiment, we vary the jammer transmit power~$\rho$ (see \fref{fig:vary_rho}). 
As expected based on the discussion in \fref{sec:barrage}, 
the performance of JASS does not vary significantly under the jammer transmit 
power (and neither does the choice of the optimal threshold parameter~$\tau$). 

In the fifth experiment, we vary the (estimated) number of jammer antennas $I$ and $\Ie$ while fixing $I=\Ie$
(see \fref{fig:vary_I}). The performance of JASS improves with decreasing number of jammer antennas. 
This is due to two reasons: The first reason is that an increased number of the interference dimensions 
increases the difficulty of the optimization problem in \eqref{eq:opt_reformulated}, which JASS solves 
approximately. The second reason is that JASS nulls $\Ie$ dimensions of the receive signal, 
which corresponds to a loss of $\Ie$ degrees of freedom (out of~$\min\{B,K\}$). 
The choice of the optimal threshold parameter~$\tau$ also varies between the scenarios, though a comparison 
with the experiment in \fref{sec:mismatch} indicates that this change is mainly due to the change in 
$\Ie$ (and not due to the change in $I$), 
which is a parameter that is known at design time of a system. 

In the final experiment, we vary the number of power iterations $t_\textnormal{max}$ that JASS uses for the eigenvector approximations
in \fref{alg:power}, and we also consider the performance of an exact eigenvalue decomposition (EVD) 
(see \fref{fig:vary_tmax}).
Naturally, the performance increases with $t_\textnormal{max}$, 
since more iterations lead to more accurate eigenvector approximations.
However, we see that already for $t_\textnormal{max}=2$ power iterations, the performance 
is virtually identical to the performance when using an exact~EVD. 

\subsection{Performance of JASS Under Different Channel Models}
\label{sec:channels}

Finally, we show that the success of JASS does not depend on the assumption of i.i.d. Rayleigh fading channels, which was made in 
\fref{sec:setup}. To this end, we consider two different scenarios in which we generate the channel vectors using the 
QuaDRiGa channel model~\cite{jaeckel2014quadriga}.

\begin{figure}[tp]
    \centering  
    \hspace{-2mm}
    \subfigure[$\rho=0$\,dB barrage jammer (\tinygraycircled{1})]{
        \includegraphics[width=0.23\textwidth]{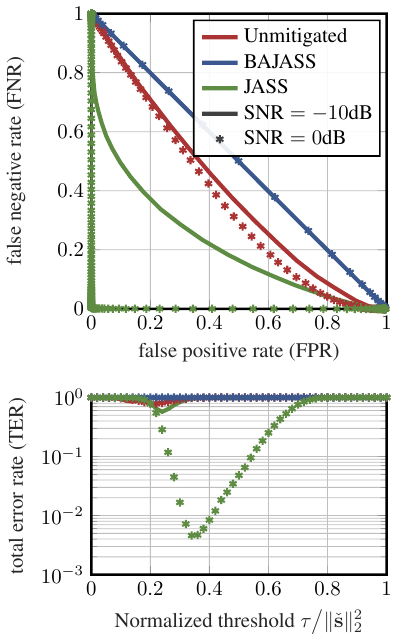}
       	\label{fig:3gpp_barrage}
    }          
    \hspace{-2mm}
    \subfigure[$\rho=30$\,dB erratic jammer (\tinygraycircled{5})]{
        \includegraphics[width=0.23\textwidth]{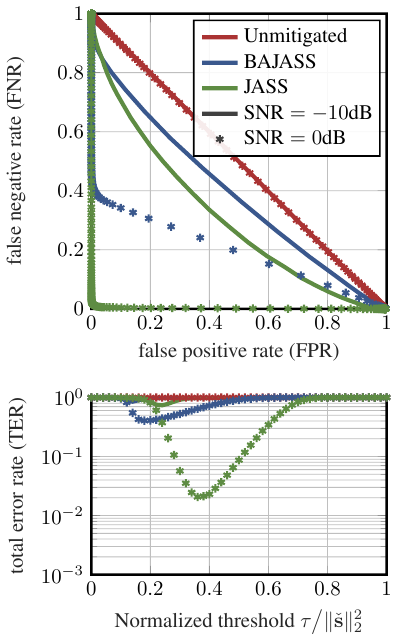}
       	\label{fig:3gpp_erratic}
    }     
    \hspace{-2mm}
    \caption{Performance on 3GPP 38.901 UMa channels with $B\!=\!16$ BS antennas.}
    \label{fig:3ggp}
\end{figure}

In the first scenario, we consider a 3GPP 38.901 urban macrocellular (UMa) channel model~\cite{3gpp22a} at a carrier
frequency of $2$\,GHz, in which the BS has a uniform linear antenna array (ULA) consisting of $B=16$ antennas spaced at 
half a wavelength. 
The UE and the jammer are randomly placed at distances between $10$\,m and $250$\,m in a $120^\circ$ angular sector 
in front of the BS, with a minimum angular separation of $1^\circ$ between them. The jammer has $I=4$ antennas, 
which are also arranged in a half-wavelength spaced ULA. 
All channels~$\bmh$ and $\bJ$ are normalized to $\|\bmh\|_2^2=B$ and $\|\bJ\|_F^2=BI$.
We exemplarily consider the mitigation of a $\rho=0$\,dB barrage jammer (\tinygraycircled{1}), as well as of an 
$\rho=30$\,dB erratic jammer (\tinygraycircled{5}).
The results are displayed in \fref{fig:3ggp}, and show that JASS is able to 
successfully mitigate both jammers also on this more realistic channel model. 
Contrasting \fref{fig:3gpp_barrage} with \fref{fig:0dbbarrage} and
\fref{fig:3gpp_erratic} with \fref{fig:erratic} shows the performance to be
extremely similar in both channel models: At $\tau=0.375\|\check\bms\|_2^2$, 
a minimum TER of slightly below $1\%$ is reached for the barrage jammer, 
and a minimum TER of around $1\%$ is reached for the erratic jammer 
(at $\text{SNR}=0$\,dB). Note that changing the channel model does not seem
to affect the choice of the optimal $\tau$. 

\begin{figure}[tp]
    \centering
    \hspace{-2mm}
    \subfigure[$\rho=0$\,dB barrage jammer (\tinygraycircled{1})]{
        \includegraphics[width=0.23\textwidth]{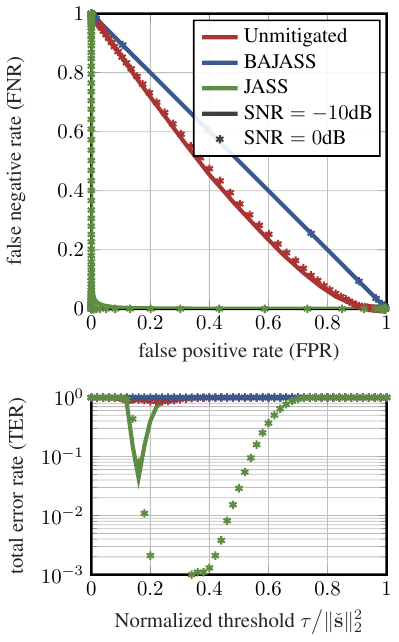}
        \label{fig:mmWave_barrage}
    }
    \hspace{-2mm}
    \subfigure[$\rho=30$\,dB erratic jammer (\tinygraycircled{5})]{
        \includegraphics[width=0.23\textwidth]{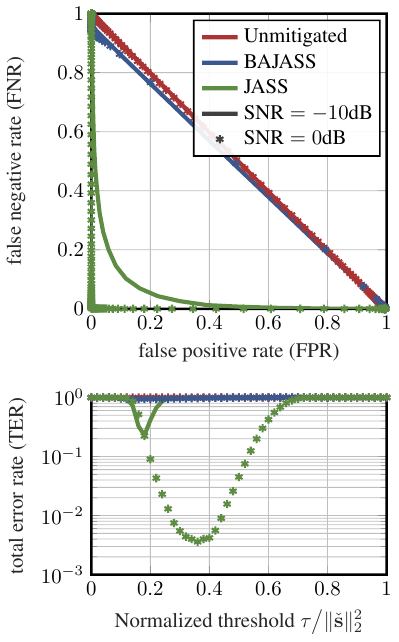}
        \label{fig:mmWave_erratic}
    }
    \hspace{-2mm}
    \caption{Performance in mmWave UMi LoS with $B=128$ BS antennas.}
    \label{fig:mmWave}
\end{figure}

In the second scenario, we consider a mmMAGIC urban microcellular (UMi) line-of-sight (LoS) model~\cite{jaeckel2014quadriga} 
at a millimeter-wave (mmWave) carrier frequency of $60$\,GHz. 
Here, the BS has $B=128$ antennas, and the distance of the UE and the jammer
ranges from $10$\,m to $100$\,m. 
All other settings are identical to the first scenario. The results are displayed in \fref{fig:mmWave}. 
Again, JASS is able to mitigate both jammers also in this channel model, which has completely different characteristics 
than Rayleigh fading. Note that, at $\text{SNR}=0$\,dB, 
$\tau=0.375\|\check\bms\|_2^2$ is still a good choice, which 
highlights the invariance of $\tau$ with respect to different channel models.
We also point out that, due to the high number of receive antennas, JASS 
even achieves fairly good performance at $\text{SNR}=-10$\,dB 
(note, however, that changes in noise power \emph{do} affect the 
choice of the optimal detection threshold~$\tau$; see \fref{fig:vary_SNR}).

\section{Conclusions}
We have proposed JASS, the first method for time-synchronization that can withstand smart jamming attacks. 
JASS mitigates jammers using spatial filtering and is based on a novel optimization problem in which 
a spatial filter is fitted to the time-windowed receive signal to maximize the normalized correlation 
with the synchronization sequence. 
We have shown the soundness of the proposed optimization problem in \fref{thm:thm}, 
where successful synchronization is guaranteed under some intuitive conditions. 
The effectiveness of the JASS algorithm itself, which solves this underlying
optimization problem only approximately, has been shown via extensive simulations
that consider different types jammers. 

The extension of our method to frequency-selective channels, as well as the 
incorporation of frequency synchronization, are left for future work.

\appendices

\section{On Distinguishing Between Small and Large Synchronization Errors}\label{app:small}

Orthogonal frequency division multiplexing (OFDM) modulation is famously able to withstand small 
synchronization errors \cite{schmidl97a}, in particular when the synchronization sequence is detected slightly too early 
(i.e., when $\hat\ell\lesssim L$). 
Nevertheless, our error metrics in \fref{sec:eval} only distinguish between whether or not \emph{perfect} synchronization $\hat\ell=L$
is achieved, but not between different error magnitudes $|L-\hat\ell|$.
In this appendix, we justify our choice of such a  binary measure of success. 

Our model assumes frequency-flat channels (cf. \fref{sec:limitations}), which are naturally combined with 
single-carrier modulation instead of OFDM. We assume frequency-flat channels not only to 
simplify the synchronization problem, but also because the combination of OFDM waveforms with MIMO jammer mitigation 
is inherently precarious because jammers may not adhere to the cyclic prefix requirement of OFDM~\cite{marti2023single}.
In single-carrier modulation, with which we envision JASS to be combined, even a slight synchronization mismatch would be fatal, 
because it leads to a misinterpretation of all subsequently transmitted symbols. 
For this reason, we do not consider it useful to distinguish between different magnitudes of synchronization errors. 

\begin{figure}
\centering
\includegraphics[width=\columnwidth]{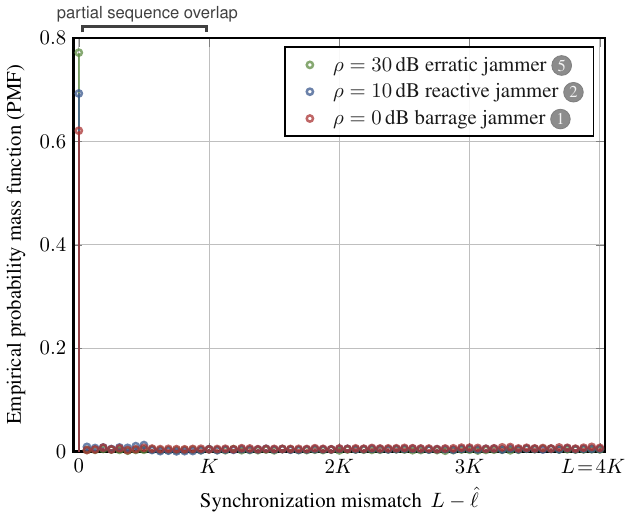}
\vspace{-3mm}	
\caption{In this experiment, we deterministically
fix the realization of $L$ to $4K=64$ in every Monte--Carlo (MC) trial, 
and we evaluate JASS for $\text{SNR}=-10$\,dB, $B=16$, 
$\Ie=I=4$, $K=16$, i.i.d. Rayleigh fading for $\bmh$ and $\bJ$, 
and $\tau=0.25\|\check\bms\|_2^2$. 
False positives and false negatives are approximately equally likely
under these parameters.
We then exclude all false negative MC trials, since we cannot evaluate 
the mismatch $L-\hat\ell$ for those; cf.~Footnote \ref{fnmismatch}. 
Among the remaining trials (i.e., those in which JASS produces the correct $\ell=L$
or a false positive $\ell<L$), we plot 
the empirical probability mass function (PMF) of the synchronization 
mismatch $L-\hat\ell$ (for different jammers). Note that the probability of 
a synchronization mismatch due to the transient behavior of a partial sequence overlap
(i.e., the mass assiociated with the discrete interval $[1,\dots,K-1\}$)
is not larger than the probability of a synchronization mismatch due to random 
noise and jamming alone (i.e., the mass assiocated with the discrete interval 
$\{K,L\}$). 
}
\label{fig:hist}
\end{figure}

Moreover, we observe empirically (see \fref{fig:hist}) that, regardless of the jammer type, 
when JASS detects the synchronization sequence too early, 
then the mismatch $L-\hat\ell$ is approximately uniformly distributed on $\{1,\dots,L\}$.\footnote{Note that we cannot investigate the distribution of $\hat\ell-L$ when JASS misses the synchronization 
sequence, since our signal model does not define what happens after; see~\eqref{eq:ue_tx}.
\label{fnmismatch}
}
This means that, when JASS incorrectly detects the sequence, it is usually not because of the transient behavior
from the partially overlapping synchronization sequence, but because of random noise. 
For this reason, it seems unnecessary to distinguish between ``small'' and ``large''
mismatches $|L-\hat\ell|$. 

\section{Proof of \fref{thm:thm}} \label{app:proof}

Let $\bA\in\opC^{B\times\Ie}$ be any matrix that satisfies $\textit{col}(\bJ)\subseteq \textit{col}(\bA)$
and $\bmh\notin\textit{col}(\bA)$. That such a matrix exists follows from Conditions 2 and 3.
For $\tilde\bP=\bI_B-\bA\herm{\bA}$ and $\ell=L$, the objective in \eqref{eq:matrix_constraint} then equals 
\begin{align}
    &\hspace{-3mm}\frac{\|\tilde\bP\bY_\ell\check \bms^\ast\|_2^2}{\|\tilde\bP\bY_\ell\|_F^2} \nonumber\\
    &\hspace{-3mm}=\frac{\|(\bI_B-\bA\herm{\bA})(\bmh\tp{\check\bms}+\bJ[\bmw[\ell],\dots,\bmw[\ell\!+\!K\!-\!1])\check \bms^\ast\|_2^2}{\|(\bI_B-\bA\herm{\bA})(\bmh\tp{\check\bms}+\bJ[\bmw[\ell],\dots,\bmw[\ell\!+\!K\!-\!1])\|_F^2}\!\! \label{eq:upper_bound_1} \\
    &\hspace{-3mm}=\frac{\left\|(\bI_B-\bA\herm{\bA})\bmh\|\check\bms\|_2^2\right\|_2^2}{\|(\bI_B-\bA\herm{\bA})\bmh\tp{\check\bms}\|_F^2}\label{eq:upper_bound_2} \\
    &\hspace{-3mm}=\frac{\left\|(\bI_B-\bA\herm{\bA})\bmh\right\|_2^2 \|\check\bms\|_2^4}{\|(\bI_B-\bA\herm{\bA})\bmh\|_2^2 \|\check\bms\|_2^2} \label{eq:upper_bound_3} \\
    &\hspace{-3mm}= \|\check\bms\|_2^2,
\end{align}
where \eqref{eq:upper_bound_2} follows from the fact that $(\bI_B-\bA\herm{\bA})\bJ=\mathbf{0}$ 
and that $\tp{\check\bms}\check\bms^\ast=\|\check\bms\|_2^2$, 
and \eqref{eq:upper_bound_3} follows from the fact that $\|\gamma\bma\|_2^2 = \gamma^2\|\bma\|_2^2$ 
and that $\|\bma\tp{\bmb}\|_F^2 = \|\bma\|_2^2 \|\bmb\|_2^2$. 

We now show that the objective in \eqref{eq:matrix_constraint} is upper-bounded by $\|\check\bms\|_2^2$,
and that (with probability one) equality cannot hold if $\ell<L$. 
From this, the result will directly follow.

That the LHS of \eqref{eq:matrix_constraint} is upper-bounded by $\|\check\bms\|_2^2$ follows~since
\begin{align}
\frac{\|\tilde\bP\bY_\ell\check \bms^\ast\|_2^2}{\|\tilde\bP\bY_\ell\|_F^2}
&\leq \frac{\|\tilde\bP\bY_\ell\|_2^2\, \|\check\bms\|_2^2}{\|\tilde\bP\bY_\ell\|_F^2} \label{eq:spectral_ineq} \\
&\leq \|\check\bms\|_2^2 \label{eq:frob_ineq},
\end{align}
where \eqref{eq:spectral_ineq} follows from the definition of the spectral norm 
and \eqref{eq:frob_ineq} follows since the spectral norm is upper bounded by the Frobenius norm.\footnote{
This follows, e.g., from the fact that the Frobenius norm of a matrix is equal to the square root 
of the sum of the squared singular values of that matrix \cite[p.\,342]{horn2013matrix}
and the fact that the spectral norm of a matrix is equal to the largest singular value of that matrix 
\cite[p.\,346]{horn2013matrix}\label{fnlabel}.}

It remains to show that, with probability one, this upper bound cannot be attained if $\ell<L$. 
The upper bound is attained if and only if both \eqref{eq:frob_ineq} and \eqref{eq:spectral_ineq} hold with equality. 
Equality in~\eqref{eq:frob_ineq} holds if and only if $\tilde\bP\bY_\ell$ is a rank-one matrix.\footnote{
This follows from Footnote \ref{fnlabel}.}
Equality in \eqref{eq:spectral_ineq} holds if and only if $\check \bms$ is collinear with the principal 
right-singular vector of $\tilde\bP\bY_\ell$ (i.e., the right-singular vector corresponding to the largest 
singular value).\footnote{This can be deduced from \cite[p.\,246]{horn2013matrix}.
If the largest singular value is not unique, 
equality holds if and only if $\check\bms$ is contained in the span of the right-singular vectors 
that correspond to the co-equal largest singular values.}
It follows that the upper bound is attained if and only if $\tilde\bP\bY_\ell$ is a rank-one matrix whose only non-degenerate
right-singular vector is collinear with $\check \bms$.
Denoting $\tp{\bms}[\ell-L] =[s[\ell-L], \dots, s[\ell-L+K-1]]$ and 
$\bW_\ell=[\bmw[\ell],\dots,\bmw[\ell+K-1]]\in\opC^{I\times K}$, we can rewrite the windowed 
receive matrix $\bY_\ell$ as
\begin{align}
    \bY_\ell = 
    \bmh \tp{\bms}[\ell-L] + \bJ\bW_\ell
    = \begin{bmatrix}
        \bmh & \bJ
    \end{bmatrix}
    \begin{bmatrix}
        \tp{\bms}[\ell-L] \\ \bW_\ell
    \end{bmatrix}. 
    \label{eq:y_prod}
\end{align}
If $\tilde\bP\bY_\ell$ is a rank-one matrix, then its compact singular value decomposition has the form
$\bmu\sigma\herm{\bmv}$, with the right-singular vector $\bmv$ satisfying $\herm{\bmv}\in\textit{row}([\tp{\bms}[\ell-L];\bW_\ell])$.

For any $k=0,\dots,K-1$ the $(k+1)$th column of $\bW_\ell$ can only depend on the first $k+1$ entries of $\tp{\bms}[\ell-L]$, 
which for $\ell<L$ cannot contain $\check{s}_{k'}, k' \geq k$. Since the entries of $\check\bms$ are i.i.d., 
it therefore follows that the $(k+1)$th column of $\bW_\ell$ is independent of $\check{s}_{k'}, k'\geq k$. 
Moreover, because the entries of $\check\bms$ are i.i.d., for $\ell<L$, the $(k+1)$th entry of 
$\tp{\bms}[\ell-L]$ is independent of~$\check{s}_{k'}, k' \geq k$.
It follows that, for $\ell<L$ and for all $k=0,\dots,K-1$, 
the $(k+1)$th column of $[\tp{\bms}[\ell-L];\bW_\ell]$ is independent of $\check{s}_{k'}, k' \geq k$.
Since the entries of $\check\bms$ are i.i.d. complex standard Gaussian, it therefore follows that, as long 
as $I+1<K$ (i.e., as long as the matrix $[\tp{\bms}[\ell-L];\bW_\ell]$ is wide), the probability that 
$\herm{\check\bms}\in\textit{row}([\tp{\bms}[\ell-L];\bW_\ell])$ is zero.

It follows that, for any fixed $\ell<L$, the probability that $\check\bms$ is collinear with $\bmv$
is zero. And since the union over a countable set of probability-zero events is zero, the probability that there 
is an $\ell<L$ for which $\check\bms$ is collinear with $\bmv$ is zero as well. 
Hence, with probability one, the upper bound $\|\check\bms\|_2^2$ cannot be attained for $\ell<L$ and the result follows. 
\hfill$\blacksquare$



\end{document}